\documentstyle[emulateapj,psfig,side]{article}
\def\gs{\mathrel{\raise0.35ex\hbox{$\scriptstyle >$}\kern-0.6em 
\lower0.40ex\hbox{{$\scriptstyle \sim$}}}}
\def\ls{\mathrel{\raise0.35ex\hbox{$\scriptstyle <$}\kern-0.6em 
\lower0.40ex\hbox{{$\scriptstyle \sim$}}}}

\slugcomment{Submitted: 1998 June 12; accepted: 1998 December 15}

\lefthead{Dressler et al.}
\righthead{A Spectroscopic Catalog of 10 Distant Rich Clusters of Galaxies}

\begin{document}

\title{A Spectroscopic Catalog of 10 Distant Rich Clusters of Galaxies}

\author{Alan Dressler,$\!$\altaffilmark{1}
Ian Smail,$\!$\altaffilmark{2,8}
Bianca M.\ Poggianti,$\!$\altaffilmark{3,6,7} 
Harvey Butcher,$\!$\altaffilmark{4}\\
Warrick J.\ Couch,$\!$\altaffilmark{5}
Richard S.\ Ellis\altaffilmark{3} \& 
Augustus Oemler Jr.\altaffilmark{1}
}

\affil{\tiny 1) The Observatories of the Carnegie Institution of Washington, 813 Santa Barbara St., 
Pasadena, CA 91101-1292}
\affil{\tiny 2) Department of Physics, University of Durham, South Rd, Durham DH1 3LE, UK}
\affil{\tiny 3) Institute of Astronomy, Madingley Rd, Cambridge CB3 OHA, UK}
\affil{\tiny 4) NFRA, PO Box 2, NL-7990, AA Dwingeloo, The Netherlands}
\affil{\tiny 5) School of Physics, University of New South Wales, Sydney 2052, Australia}
\affil{\tiny 6) Royal Greenwich Observatory, Madingley Rd, Cambridge CB3 0EZ, UK}
\affil{\tiny 7) Osservatorio Astronomico di Padova, vicolo dell'Osservatorio 5, 35122 Padova, Italy}

\altaffiltext{8}{Visiting Research Associate at the Carnegie
Observatories.}
\setcounter{footnote}{8}

\begin{abstract}
We present spectroscopic observations of galaxies in the fields of 10
distant clusters for which we have previously presented deep imaging
with {\it WFPC2} on board the {\it Hubble Space Telescope}.  The clusters
span the redshift range $z=0.37$--0.56 and are the subject of a
detailed ground- and space-based study to investigate the evolution of
galaxies as a function of environment and epoch.  The data presented
here include positions, photometry, redshifts, spectral line strengths
and classifications for 657 galaxies in the fields of the 10 clusters.
The catalog comprises 424 cluster members across the 10 clusters and 233
field galaxies, with detailed morphological information from our {\it WFPC2}
images for 204 of the cluster galaxies and 71 in the field.  We
illustrate some basic properties of the catalog, including correlations
between the morphological and spectral properties of our large sample
of cluster galaxies.  A direct comparison of the spectral properties of
the high redshift cluster and field populations suggest that the
phenomenon of strong Balmer lines in otherwise passive galaxies
(commonly called E+A, but renamed here as the k+a class) shows an
order-of-magnitude increase in the rich cluster environment, compared
to a more modest increase in the field population. This suggests that
the process or processes involved in producing k+a galaxies are either
substantially more effective in the cluster environment or that this
environment prolongs the visibility of this phase. A more detailed
analysis and modeling of these data will be presented in Poggianti et
al.\ (1998).
\end{abstract}

\keywords{galaxies: clusters: general -- galaxies: evolution}

\section{Introduction}

%
%
\begin{table*}
{\scriptsize
\begin{center}
\centerline{\sc Table 1}
\vspace{0.1cm}
\centerline{\sc Telescope and Instrument Log}
\vspace{0.3cm}
\begin{tabular}{lcccccl}
\hline\hline
\noalign{\smallskip}
 {Telescope} & {Instrument} & {$\lambda$} & {Spectral Scale} & {Spatial Scale} & \# Nights & {Reference} \cr
~ & ~ & (\AA) & (\AA/px) & (arcsec/px) & \cr
\hline
\noalign{\smallskip}
Palomar 5.1-m & COSMIC & 3500--9800 & 3.1 & 0.40 & 19 & Kells et al.\ (1998) \cr
WHT 4.2-m & LDSS-2 & 3500--8300 & 5.3 & 0.59 & ~6 & Allington-Smith et al.\ (1994)\cr
NTT 3.5-m & EMMI & 3600--7800 & 2.3 & 0.27 & ~2 & Zijlstra et al.\ (1996) \cr
\noalign{\smallskip}
\noalign{\hrule}
\noalign{\smallskip}
\end{tabular}
\end{center}
}
\vspace*{-0.8cm}
\end{table*}

The change with redshift observed in the proportion of star-forming
galaxies in the cores of rich clusters was uncovered over twenty years
ago, by Butcher \& Oemler (BO, 1978, 1984), but it remains one of the
clearest and most striking examples of galaxy evolution.  Considerable
effort has gone into acquiring photometric information that would
elucidate the physical processes active in distant clusters and their
effects on the evolution of both the star-forming (Lavery \& Henry
1994; Lubin 1996; Rakos \& Schombert 1995; Rakos, Odell \& Schombert
1997) and passive galaxies (Arag\'on-Salamanca et al.\ 1993; Stanford,
Eisenhardt \& Dickinson 1995, 1998; Smail et al.\ 1998).  Further
impetus has been provided by observations of the recent transformation
of the S0 population of clusters (Dressler et al.\ 1997), which may
allow a closer connection to be drawn between the galaxy populations of
distant clusters and the evolutionary signatures found in their local
Universe counterparts (Caldwell \& Rose 1997; Bothun \& Gregg 1990).

However, it was the advent of spectroscopic surveys of the distant
cluster populations (e.g.\  Dressler \& Gunn 1983, 1992, DG92; Couch \&
Sharples 1987, CS87; Barger et al.\ 1996; Abraham et al.\ 1996; Fisher
et al.\ 1998) which uncovered the real breadth of the changes in
galaxies in these environments, including several spectral signatures
of evolutionary change, such as evidence for a strong decline in the
star-formation rates of many cluster galaxies in the recent past.  The
advent of  high spatial resolution imaging with the {\it Hubble Space
Telescope} ({\it HST}\,) provided a further breakthrough, giving
morphological information on the galaxies in these distant clusters.
This could be used to link the evolution of stellar populations in the
galaxies with the evolution of their {\it structure}, in order to
understand how the various galaxy types we see in the local universe
came to be.  Pre- and Post-refurbishment {\it HST} observations by two
groups (Couch et al.\ 1994, 1998; Dressler et al.\ 1994; Oemler et
al.\ 1997) were used in early attempts to correlate spectral evolution
with morphological/structural data, and to provide some insight into
the mechanisms that might be driving the strong evolution in the
cluster galaxy population.  These two programs were extended from
Cycle-4 into the ``MORPHS'' project, which accumulated
post-refurbishment {\it WFPC2} images for 11 fields in 10 clusters at $z =
0.37$--0.56, viewed at a time some 2--4\,$h^{-1}$ billion years before
the present day.\footnote{We use $q_\circ=0.5$ and $h=1$, where $h =
H_\circ / 100$ km s$^{-1}$ Mpc$^{-1}$.  For this geometry 1 arcsec is
equivalent to 3.09\,$h^{-1}$ kpc for our lowest redshift cluster and
3.76\,$h^{-1}$ kpc for the most distant.} The photometric and
morphological galaxy catalogs from these images were presented in Smail
et al.\ (1997b, S97), while the data have also been used to study the
evolution of the early-type galaxies within the clusters, using both
color (Ellis et al.\ 1997) and structural information (Barger et
al.\ 1998), the evolution of the morphology-density relation of the
clusters (Dressler et al.\ 1997) and the masses of the clusters from
weak lensing (Smail et al.\ 1997a).

The aim of this paper is to combine the morphological information
available from our {\it HST} images with detailed star-formation
properties of the cluster galaxies derived from targeted spectroscopic
observations.  To this end we have used over 27 clear, dark nights over
the past 4 years on the Palomar 5.1-m (P200),\footnote{The Hale 5-m of the
Palomar Observatory is owned and operated by the California Institute
of Technology.} 4.2-m William Herschel Telescope (WHT)\footnote{The
William Herschel Telescope of the Observatorio del Roques de los
Muchachos, La Palma, is operated by the Royal Greenwich Observatory on
behalf of the UK Particle Physics and Astronomy Council.} and the 3.5-m
New Technology Telescope (NTT)\footnote{Based in part on observations
collected at the European Southern Observatory, La Silla, Chile.} to
assemble a large catalog of spectroscopic data on galaxies in these
clusters.  We combine these new observations with previously published
spectroscopy from DG92 and present spectroscopic observations of a
total of 424 cluster members, of which 204 have morphologies from our
{\it HST} imaging, as well as 233 field galaxies (71 with {\it HST}
morphologies).  In addition, we have analyzed all of the spectra to
provide equivalent width measurements on a uniform system for the
entire sample.  The spectral catalogs, including line strength and
color information, as well as the reduced spectra themselves in {\sc
FITS} format, are available at the AAS web site.  A more detailed
analysis of the spectroscopic data presented here will be given in
Poggianti et al.\ (1998, P98).

A plan of the paper follows. We start by discussing the observations
and their reduction in \S2.  In \S3 we then give the details of the
redshift measurements, as well as our analysis to quantify the
strengths of spectral features and information about our spectral
classification scheme based upon these.  We then present the spectral
properties of galaxies in the catalog and relate these to the
morphologies of the galaxies from our {\it HST} images in \S4, before
discussing our results in \S5.  Finally in \S6 we list the main
conclusions of this work.

\section{Observations and Reduction}

%
%
\begin{table*}

{\scriptsize
\begin{center}
\centerline{\sc Table 2}
\vspace{0.1cm}
\centerline{\sc Properties of The Clusters}
\vspace{0.3cm}
\hspace*{-0.5cm}\begin{tabular}{lcccccll}
\hline\hline
\noalign{\smallskip}
 {Cluster} & {z} & {$\sigma_{\rm cl}$} & {$\Delta$z} &  R.A. & Dec. &
 \hfil {\it HST} Filters & Comment \cr
~ & ~ & (km s$^{-1}$) & ~ & (J2000) & (J2000) & ~ &\cr
\hline
\noalign{\smallskip}
A\,370        & 0.3741 & 1170 & 0.3589--0.3873 &  02~39~52.6 & $-$01~34~18 & F555W/F814W, F702W & 
Center (F702W), outer field (F555W/F814W) \cr
Cl\,1447$+$26 & 0.3762 & 1470 & 0.3621--0.3857 &  14~49~29.3 & $+$26~07~52 & F702W & \cr
Cl\,0024$+$16 & 0.3928 & 1150 & 0.3755--0.4081 &  00~26~35.7 & $+$17~09~46 & F450W/F814W & \cr
Cl\,0939$+$47 & 0.4060 & 1260 & 0.3879--0.4173 &  09~42~56.1 & $+$46~59~12 & F555W/F814W, F702W & 
Center (F702W), outer field (F555W/F814W) \cr
Cl\,0303$+$17 & 0.4184 & 1310 & 0.4018--0.4338 &  03~06~12.9 & $+$17~20~08 & F702W & \cr
3C\,295       & 0.4593 & 1630 & 0.4464--0.4733 &  14~11~10.5 & $+$52~12~11 & F702W & \cr
Cl\,0412$-$65 & 0.5074 & ~700 & 0.5024--0.5130 &  04~12~50.1 & $-$65~50~44 & F555W/F814W & \cr
Cl\,1601$+$42 & 0.5388 & 1210 & 0.5100--0.5473 &  16~03~12.0 & $+$42~45~26 & F702W & \cr
Cl\,0016$+$16 & 0.5459 & 1660 & 0.5300--0.5601 &  00~18~33.3 & $+$16~26~16 & F555W/F814W & Spectra from DG92 \cr
Cl\,0054$-$27 & 0.5608 & 1180 & 0.5520--0.5770 &  00~56~59.0 & $-$27~40~20 & F555W/F814W & \cr

\noalign{\smallskip}
\noalign{\hrule}
\noalign{\smallskip}
\end{tabular}
\end{center}
}
\vspace*{-0.8cm}
\end{table*}

\subsection{Selection of Spectroscopic Targets}

The new spectroscopic observations discussed here were targeted at
determining the membership of the numerous distorted and irregular
galaxies revealed by our {\it HST} {\it WFPC2} images of the clusters, as
well as gaining a more complete understanding of the star-formation
properties of the general cluster population.  With these aims, the
object selection is closer to that employed by DG92, than the
magnitude-limited selection criteria of CS87 and Barger et
al.\ (1996).  The latter approach has some claim to making the
subsequent analysis simpler, especially when the sample is selected in
the near-IR.  However, it is a very inefficient method for studying the
faint, blue cluster members as it produces samples dominated by passive
spheroidal cluster members.  We chose instead to base our object selection
upon galaxy morphology within the region covered by our {\it WFPC2} imaging,
while being approximately magnitude-limited outside that area
(selected from ground-based $r$ or $i$ CCD material to limits of $r\sim
22$ and $i\sim 21$).  We note at this point that two of the cluster
fields, A\,370 Field\,2 and Cl\,0939$+$47 Field\,2, lie outside of the
central regions of their respective clusters (although we do also have
observations of the core regions as well).  The difference in the
galaxy density between the fields should be kept in mind in the
following analysis, although we will highlight such selection effects
for individual figures when they are discussed below.  Modelling of the
sample selection for the entire spectroscopic catalog is dealt with in
more detail in P98.

\subsection{Spectroscopic Observations}

The spectroscopic observations discussed in this paper were undertaken
with a variety of facilities over the period 1993--1997.  We list the
instruments and telescopes employed and the total number of nights used
in Table~1.  The basic details of the 10 clusters targeted in this
study are listed in Table~2, this includes the mean cluster redshift,
the one dimensional velocity dispersion ($\sigma_{cl}$, see \S3.2), the redshift
range used to define cluster membership ($\Delta z$), the field center
and the {\it HST} {\it WFPC2} filters used in the observations. The new spectra
presented here are typically of high quality due to both the long
exposure times employed in our observations and the combination of the
high efficiency of the multi-object spectrographs and the large
aperture of the telescopes used. We give in Table~3 the logs of the
observing runs for the various telescopes.  We list the mask
identification, the dates of the observations, the total exposure time and
the number of objects extracted from each mask (N).   The slit width
typically used was 1.5\,arcsec, with slits between 10--20 arcsec long.
The exact size of the region on the slit used to extract the galaxy
spectrum depended upon the relative signal to noise of the galaxy
spectrum, but varied between 1.1--8.4\,arcsec for the COSMIC spectra
with a mean length of $3.9\pm 1.2$\,arcsec.  At the median redshift of
the clusters in our catalog, the spectra thus sample a physical scale
of $\sim (5 \times 13) h^{-1}$\, kpc.

The exact details of the extraction and reduction of the spectra
depends upon the instrument and set-up used.  However, the basic steps
were the same for all the data and we outline the procedures used for
both the COSMIC and WHT/NTT data.  The raw frames were debiased using
the over-scan regions on the chip, before being trimmed.  A two
dimensional flatfield was constructed by dividing the flatfield
exposure by a low-order fit in the dispersion direction.  The data
frame was then divided by this normalized flatfield, this served to
correct for the pixel-to-pixel response of the detector.  The sequence
of data frames for each mask taken on a single night were then checked
for spatial offsets between the exposures arising from flexure in the
spectrograph (these are typically only $\ls 0.2$ pixels for COSMIC in
the course of a night).  If necessary the exposures were shifted in the
spatial and/or dispersion direction to align them and then combined
with a cosmic-ray rejection algorithm using the {\sc IRAF} task {\sc
imcombine}.  This produced a two dimensional image of the mask exposure
clean of cosmic ray events.  These frames were then geometrically
remapped to align the spectra along the rows of the detector.  This
step is necessary to remove the distortion of the spectra on the
detector introduced by the spectrograph optics. The distortion is only
a large effect for objects in slits near the edge of COSMIC's large
$13.7' \times 13.7'$ field of view, although aligning the spectra also
helps when tracing some of the faintest objects.  The distortion of the
spectra are mapped using the positions of the emission lines in the arc
exposure taken after every science exposure.  The positions of objects
in each slit on the remapped frame, as well as regions of clear sky
surrounding them, were then defined interactively using the {\sc IRAF}
package {\sc apextract}.  The exact position of the object within the
slit was traced in the dispersion direction and fitted with a low-order
polynomial to allow for atmospheric refraction.  The spectra
were then sky-subtracted and extracted using optimal weighting to
produce one dimensional spectra.  The arc exposures associated with
each science exposure were remapped and  extracted in exactly the same
manner (although with no sky-subtraction) and these were used to
determine the wavelength calibration for the science exposure.  We
estimate our wavelength scale is good to 0.2\AA\ rms.  Finally, the one
dimensional spectra were smoothed to the instrumental resolution, $\sim
8$\AA, and rebinned to 10\AA\ per pixel to make them more manageable.
The spectra obtained with COSMIC have not been flux calibrated.

The WHT and NTT spectra have been reduced using the {\sc LEXT} package,
purposely written for reducing LDSS--2 spectra, and the {\sc MIDAS}
software package.  What follows is a brief description of the reduction
procedure generally adopted. A number of twilight and dome flatfields,
and several arc frames were obtained for each mask, as well as numerous
bias frames and long--slit spectra of standard stars for flux
calibration (at least one star per night).  The raw frames were first
debiased and then divided by the corresponding normalized flatfield.
They were then calibrated in wavelength with the arcs frames obtained
either with a CuAr or HeArNe combination of lamps. The sky--subtraction
step was performed with an interactive choice of the spatial limits of
the spectrum, which was then extracted summing the counts weighted with
a Gaussian.  The long--slit stellar spectrum was reduced in a similar
way as the target spectra and a response function was derived by the
comparison with a tabulated spectrum. Each spectrum was
flux--calibrated in $F_{\nu}$ by dividing for this response function.
In the case of the WHT and NTT spectra each exposure of a given mask
was reduced and calibrated separately, before all the spectra of a
given galaxy were coadded; no smoothing or rebinning was applied.

The full digital catalog of {\sc FITS} spectra collected for this program is
distributed in electronic form on the AAS web site.  These
spectra are also available from:  http://www.ociw.edu/$\sim$irs.

%
%
\begin{table*}
\vspace*{-0.5cm}
{\scriptsize
\begin{center}
\centerline{\sc Table 3}
\vspace{0.1cm}
\centerline{\sc Log of New Spectroscopic Observations}
\vspace{0.3cm}
\hspace*{-1.3cm}\begin{tabular}{llcrc}
\hline\hline
\noalign{\smallskip}
 {Target/Mask} &  {~~~~~~~~~~~~~~Date} & 
  {Exposure time} & {N} & {Comments} \cr
~ &   &
{(ks)} &  & \cr
\hline
\noalign{\smallskip}
\multispan5{\hfil COSMIC/P200\hfil }\cr
\noalign{\smallskip}
Cl\,0024$+$16 EW--1 & 5--6 Aug 1994 &  18.8 & 37 & \cr
Cl\,0024$+$16 EW--2 &  29 Nov 1994, 1--3 Dec 1994 &  20.8 & 34 & \cr
Cl\,0024$+$16 NS--1 &  29--30 Oct 1994 &  27.0 & 35 & \cr
Cl\,0024$+$16 NS--2 &  19 Aug 1995, 25--26 Sept 1995 &  20.0 & 29 & \cr
\noalign{\smallskip}
A370 MS--1 & 25--26 Sept 1995 &  10.8 & 25 & \cr
A370 MS--2 & 27--28 Nov 1995 &  24.0 & 29 & \cr
A370 MS--3 & 4 Oct 1997 & 16.2 & 22 & \cr
\noalign{\smallskip}
Cl\,0303$+$17 EW--1 &  29 Nov 1994, 1 Dec 1994 &  25.0 & 30 & \cr
Cl\,0303$+$17 EW--2 &  2 Dec 1994 &  15.0 & 30 & \cr
Cl\,0303$+$17 NS--1 &  3 Dec 1994 &  15.0 & 27 & \cr
\noalign{\smallskip}
Cl\,0939$+$47 NS--B &  13 Dec 1993 &  10.8 & 18 & Using `old' COSMIC CCD\cr
Cl\,0939$+$47 EW--B & 29--30 Nov 1994, 1 Dec 1994 &  38.2 & 35 & \cr
Cl\,0939$+$47 EW--F & 1--2 Dec 1994, 24 Feb 1995 &  36.0 & 34 & \cr
\noalign{\smallskip}
Cl\,1601$+$42 CM--1 &  28--29 May 1997 &  22.2 & 20 & \cr
Cl\,1601$+$42 CM--2 & 18--19 June 1996 &  21.6 & 22 & \cr
Cl\,1601$+$42 CM--3 & 11--12 May 1997, 29 May 1997 &  30.2 & 24 & \cr
\noalign{\bigskip}
\multispan5{\hfil LDSS-2/WHT\hfil }\cr
\noalign{\smallskip}
Cl\,0939$+$47 WA--1 &  24 Mar 1993 & 14.4 & 10 & \cr
Cl\,0939$+$47 WA--2 & 25 Mar 1993 & 14.4 &  7 & \cr
Cl\,0939$+$47 WA--3 & 26 Mar 1993 & 10.8 & 7 & \cr
Cl\,0939$+$47 WA--4 & 26 Mar 1993 & ~9.0  & 6 & \cr
Cl\,0939$+$47 MA--A,MA--D & 27--28 Apr 1995 &  ~9.0 & 10 & \cr
Cl\,0939$+$47 MB--A,MB--D & 29 Apr 1995 &  ~9.0 & 7 & \cr
\noalign{\smallskip}
3C\,295 MA--A,MA--D &  27 Apr 1995 & ~9.0 & 6 & \cr
\cr
\noalign{\smallskip}
Cl\,1447$+$26 WA-1 & 25-26 Mar 1993 & 10.8 & 23 & \cr
Cl\,1447$+$26 MA-A,MA-D & 28--29 Apr 1995 & 12.6 & 9 & \cr
\noalign{\smallskip}
Cl\,1601$+$42 MA-A,MA-D & 28 Apr 1995 & ~9.0  & 8 & \cr
\noalign{\bigskip}
\multispan5{\hfil EMMI/NTT\hfil }\cr
\noalign{\smallskip}
Cl\,0054$-$27 MA-1 & 23 Nov 1995 &  10.2 & 15 & \cr
Cl\,0054$-$27 MA-2 & 24 Nov 1995 &  10.8 & 21 & \cr
\noalign{\smallskip}
Cl\,0412$-$65 MA-1 & 23 Nov 1995 & ~9.0 & 15 & \cr
Cl\,0412$-$65 MA-2 & 24 Nov 1995 &  8.6 & 21 & \cr
\noalign{\smallskip}
\noalign{\hrule}
\noalign{\smallskip}
\end{tabular}
\end{center}
}
\vspace*{-0.8cm}
\end{table*}

\section{Spectroscopic Analysis}

The full catalog of objects observed spectroscopically in the 10
clusters is given in Tables~4 (the complete tables are included on the
AAS web site as well as being available from
http://www.ociw.edu/$\sim$irs).  This has been split into ``Cluster''
and ``Field'' samples as described below.  The tables list not only the
spectral information on the galaxies, but also any  available
morphological and photometric data from S97 and DG92.  A key to the
various parameters and the format of the tables are given in Table~5.
We now describe in more detail the measurement of some of the spectral
parameters listed in Tables~4.

\subsection{Spectral Measurements}

The quality of the spectra, both in terms of signal-to-noise and
sky-subtraction, was visually assessed by AD for {\it all} of the
spectra presented.  The spectra are graded on a 4--point range, with
$q=1$ signifying the best and $q=4$ the worst quality.  Of the complete
catalog 17\% have $q=1$, 47\% with $q\leq 2$ and 89\% are $q\leq 3$.
Spectra with $q\leq 3$ have sufficient signal to noise (S/N) for not
only measurement of a redshift, but also to quantify the strength of
any spectral features present.  ~From the continuum regions around the
[O{\sc ii}]$\lambda$3727 and H$\delta$ lines we estimate median S/N of
40.2 ($q=1$), 28.3 ($q=2$) and 19.7 ($q=3$), with lower limits to the
S/N of 20.9, 10.6 and 4.6 respectively for these three quality classes.
Repeated observations suggest that the redshifts of $q=1$ and $q=2$
cases are correct at a confidence of greater than 98\%, and that $q=3$
cases are correct at a confidence of greater than 90\%. In contrast,
those spectra with $q=4$ are of sufficient S/N to provide only a
redshift, which may be uncertain in a significant number of cases.

Redshifts were measured from the spectra interactively using
purpose-written software that compares the wavelengths of redshifted
absorption and emission lines with features in the spectra.
Whenever possible we used a number of features to estimate the
redshift, and only in a very small number of cases is a redshift
based on only a single feature --- these instances are noted in the
comments in Tables~4.  We list in column 24 of Tables~4 the main
features used to identify the galaxy redshifts.  For conciseness we
have used the following abbreviations to identify the lines:  Babs,
Balmer absorption lines; Ha, H$\alpha$; Hb, H$\beta$; Hd, H$\delta$;
He, H$\epsilon$; Heta, H$\eta$; Hg, H$\gamma$; Hth, H$\theta$; Hz,
H$\zeta$; G, G-band; H\&K, Ca H or K; Mg, Mg-B; Na, Na-D; OII, [O{\sc
ii}]$\lambda$3727; OIII, [O{\sc iii}]$\lambda$4959,5007; bk, 4000\AA\
break; MgII, Mg{\sc ii}$\lambda$2799; CIII, C{\sc iii}]$\lambda$1909;
CIV, C{\sc iv}$\lambda$1549; FeI, Fe{\sc i}$\lambda$5268; NII, [N{\sc
ii}]$\lambda$6583; SII, [S{\sc ii}]$\lambda$6716,6731.  

The strength of emission and absorption features in the spectra were
measured using purpose-written software, allowing the positioning of
the continuum to be defined interactively.  We give the restframe
equivalent widths (EW) for [O{\sc ii}]$\lambda$3727 and H$\delta$ in
columns 5 and 6 of Table~4A and 4B, in all instances a line seen in
emission is given a negative value and is quoted in \AA.   The presence
and strength of these lines is used in the spectral classification
scheme discussed in \S 3.2.  If other lines in the spectrum were
measurable we list their EW in the comments.   We give line strengths
for not only those galaxies  observed for this work, but also those
from the early survey of DG92.  The D4000 measurements have been
similarly placed on a consistent system. These are measured using
wavelength intervals as defined in Dressler and Shectman (1987). The
COSMIC data shared a common relation of counts to flux, but were not
flux calibrated per se.  A multiplicative correction of 1.34 to convert
the measured D4000 to true D4000 for these data was derived by
comparing the COSMIC spectra of repeated objects with the equivalent
flux calibrated DG92 spectra.  This procedure, though imperfect,
generates reasonable and consistent results, as shown by multiple
COSMIC observations of the same galaxies.

We have a total of 31 repeat observations, both internally within the
datasets from a single telescope, and between telescopes.  We find
median rms scatters of $\sigma(z_{\rm COSMIC} - z_{\rm DG92}) = 0.0018$
($N=14$), $\sigma(z_{\rm COSMIC} - z_{\rm WHT}) = 0.0009$ ($N=2$) and
$\sigma(z_{\rm COSMIC} - z_{\rm COSMIC}) = 0.0005$ ($N=7$) for those
spectra with $q\leq 3$, and no systematic offsets between any of the
individual datasets:  $<\! z_{\rm COSMIC} - z_{\rm DG92}\! > = 0.0007$,
$<\! z_{\rm COSMIC} - z_{\rm WHT}\! > = -0.0009$.  We therefore
conclude that there are no significant offsets between the redshifts
from the different datasets and hence we are confident that we can
include all the  observed objects in our analysis.

%
%
\begin{table*}
{\scriptsize
\begin{center}
\centerline{\sc Table 5}
\vspace{0.1cm}
\centerline{\sc Notes on The Parameters in Tables~4}
\vspace{0.3cm}
\begin{tabular}{lllll}
\hline\hline
\noalign{\smallskip}
 {Column} &  ~~~{Parameter} & {Units} & {Format} & ~~~~~~~~{Comment} \cr
\hline
\noalign{\smallskip}
 ~1..... & CLUSTER       &        & A6    & Cluster \cr
 ~2..... & ID            &        & I4    & ID in spectroscopic catalog for cluster \cr
 ~3..... & $z$           &        & F7.4  & Redshift\cr
         &               &        & A1    & Redshift quality -- a ``:'' indicates questionable identification \cr
 ~4..... & $Q$           &        & I2    & Quality of spectrum: 1=High, 4=Low. \cr
 ~5..... & [O{\sc ii}]   & \AA    & F7.1  & Restframe equivalent width of [O{\sc ii}]\,3727\cr
         &               &        & A1    & Quality of [O{\sc ii}]\,3727 EW measurement (a ``:'' indicates questionable)\cr
 ~6..... & H$\delta$     & \AA    & F4.1  & Restframe equivalent width of H$\delta$, $-$ve indicates emission  \cr
         &               &        & A1    & Quality of H$\delta$ EW measurement (a ``:'' indicates questionable)\cr
 ~7..... & D4000         &        & F5.2  & Break strength index \cr
 ~8..... & CLASS         &        & A11   & Spectral classification in scheme described in \S 3.3\cr
 ~9..... & $\delta$RA    & arcsec & F7.1  & RA offset from field center in Table~2 \cr
 10..... & $\delta$Dec   & arcsec & F7.1  & Dec offset from field center in Table~2 \cr
 11..... & ID$_{HST}$    &        & I5    & ID in photometric catalog for cluster$^a$ \cr
 12..... & X             & Pixels & I5    & X coordinate on {\it WFPC2} frame$^a$ \cr
 13..... & Y             & Pixels & I5    & Y coordinate on {\it WFPC2} frame$^a$ \cr
 14..... & MORPH         &        & A12   & Galaxy morphology$^a$ \cr
 15..... & T             &        & I2    & T-type$^a$ \cr
 16..... & D             &        & I2    & Visual disturbance index$^a$ \cr
 17..... & INT           &        & A6    & Interpretation of disturbance$^a$ \cr
 18..... & MAG           & Mag    & F5.2  & Total magnitude in F702W/F814W from {\it WFPC2} frame$^{a,b}$ \cr
 19..... & COL           & Mag    & F5.2  & Aperture color from {\it WFPC2} frame$^{a,c}$\cr
 20..... & MAG$_{\rm DG}$ & Mag   & F6.2  & Magnitude from ground-based imaging published in DG92$^d$ \cr
 21..... & COL$_{\rm DG}$ & Mag   & F6.2  & Color from ground-based imaging published in DG92$^d$ \cr
 22..... & RUN           &        & A6    & Code giving details of observing run$^e$  \cr
 23..... & MASK          &        & A10   & Mask and object slit identifier \cr
 24..... & FEATURES      &        & A23   & Spectral features identified, see \S 3.1\cr
 25..... & COMMENTS      &        & A130  & Description of features in spectrum \cr
\noalign{\smallskip}
\noalign{\hrule}
\noalign{\smallskip}
\multispan5{~~~$^a$ See S97 for more details. \hfil}\cr
\noalign{\vspace{0.1cm}}
\multispan5{~~~$^b$ Magnitudes are in F702W for Cl\,0303$+$17, Cl\,0939$+$47, 
3C\,295, Cl\,1447$+$26 and Cl\,1601$+$42, and in F814W \hfil}\cr
\noalign{\vspace{0.05cm}}
\multispan5{ 
for Cl\,0016$+$16, Cl\,0024$+$16, Cl\,0054$-$27, A\,370 Field\,2, Cl\,0412$-$65,
Cl\,0939$+$47 Field\,2. \hfil}\cr
\noalign{\vspace{0.1cm}}
\multispan5{~~~$^c$ {\it WFPC2} $V_{555}-I_{814}$ color information is available for: Cl\,0016$+$16, Cl\,0054$-$27, 
A\,370 Field\,2, Cl\,0412$-$65,  Cl\,0939$+$47, \hfil}\cr
\noalign{\vspace{0.05cm}}
\multispan5{and $B_{450}-I_{814}$ colors for Cl\,0024$+$16. \hfil}\cr
\noalign{\vspace{0.1cm}}
\multispan5{~~~$^d$ Aperture $r$-band magnitude from DG92, colors are 
aperture $(g-r)$ measurements in all instances.  \hfil}\cr
\noalign{\vspace{0.1cm}}
\multispan5{~~~$^e$ [P/W/N]$<$MONTH$>$$<$YEAR$>$, P=Palomar 5-m, W=WHT, N=NTT, or DG92. \hfil}\cr
\end{tabular}
\end{center}
}
\vspace*{-0.8cm}
\end{table*}

Finally, we quantified the detectability of [O{\sc ii}] and H$\delta$
in our spectra.  This enabled us to derive the lower limits on the
strength of these spectral features below which we would not have
identified them. Achieving this aim was not straightforward because the
code that best measured the equivalent widths, which is based on a
gaussian line-fitting program written by Paul Schechter, does not
perform well when the lines are weak or undetectable.  For this reason,
when we measured the  strengths of features in those galaxies where the
feature was not clearly seen, we by necessity had to measure equivalent
widths using the standard technique of obtaining the continuum level
from straddling continuum bands, and measuring the decrement or
increment in signal relative to the continuum in an interval containing
the feature.  We made such measurements of [O{\sc ii}] and H$\delta$ EW
for all COSMIC spectra with qualities $q\leq 3$ of cluster members in
Cl\,0939$+$47 and Cl\,0024$+$16, a total of 79 galaxies. The intervals
are, again, as defined in Dressler and Shectman (1987).  For weak, but
measurable, cases the line-fitting and flux-summing techniques give
equivalent results, though for strong absorption lines, in particular,
the latter seems to underestimate the strength of the feature,
apparently by allowing the wings of the line to lower the continuum
level.  We believe, however, that the two scales for measuring
equivalent widths are interchangeable for the purpose of looking for
weak features.

%
%
\begin{figure*}
\centerline{\psfig{file=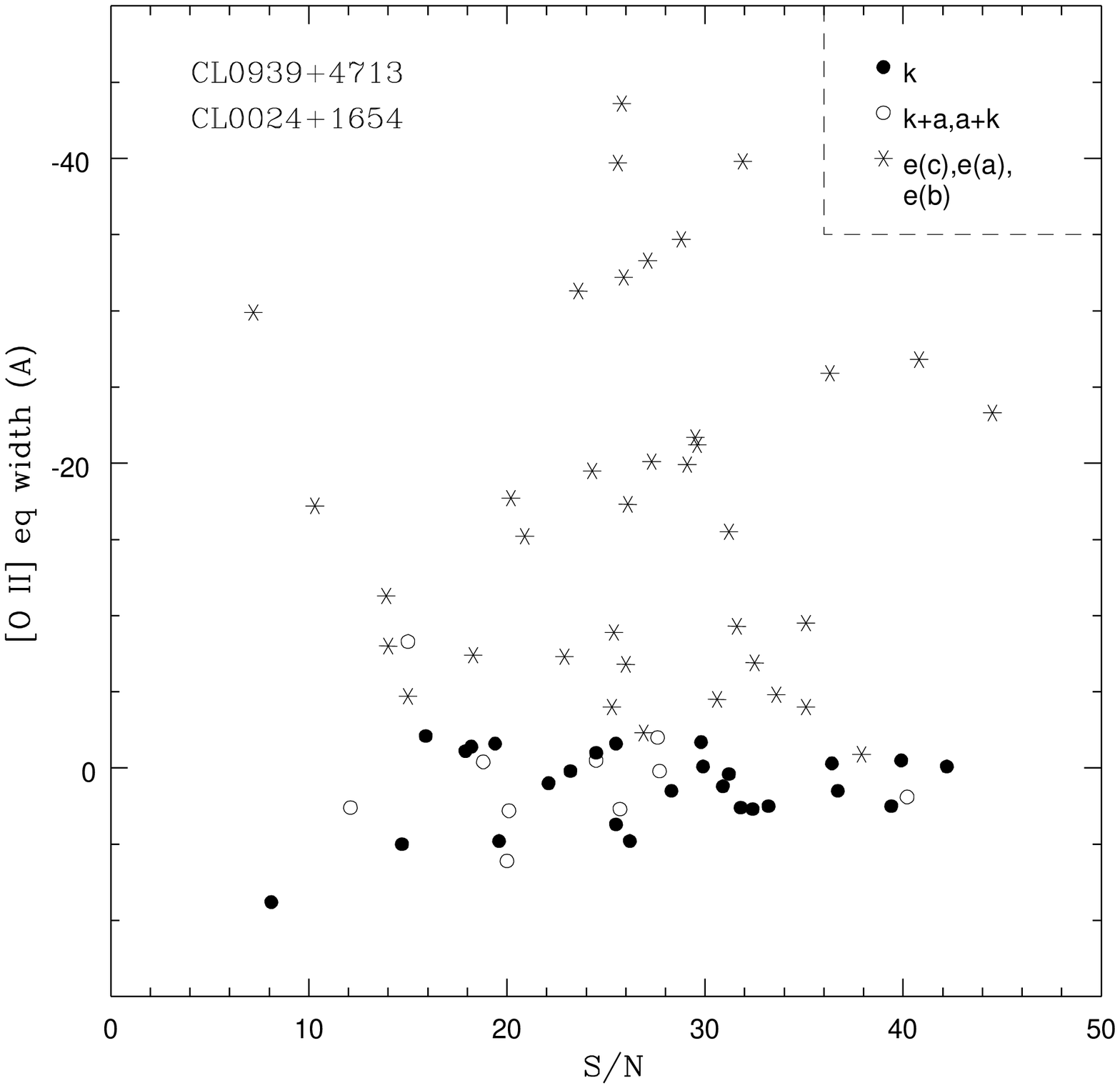,angle=0,width=3.0in}\hspace*{0.5cm}
\psfig{file=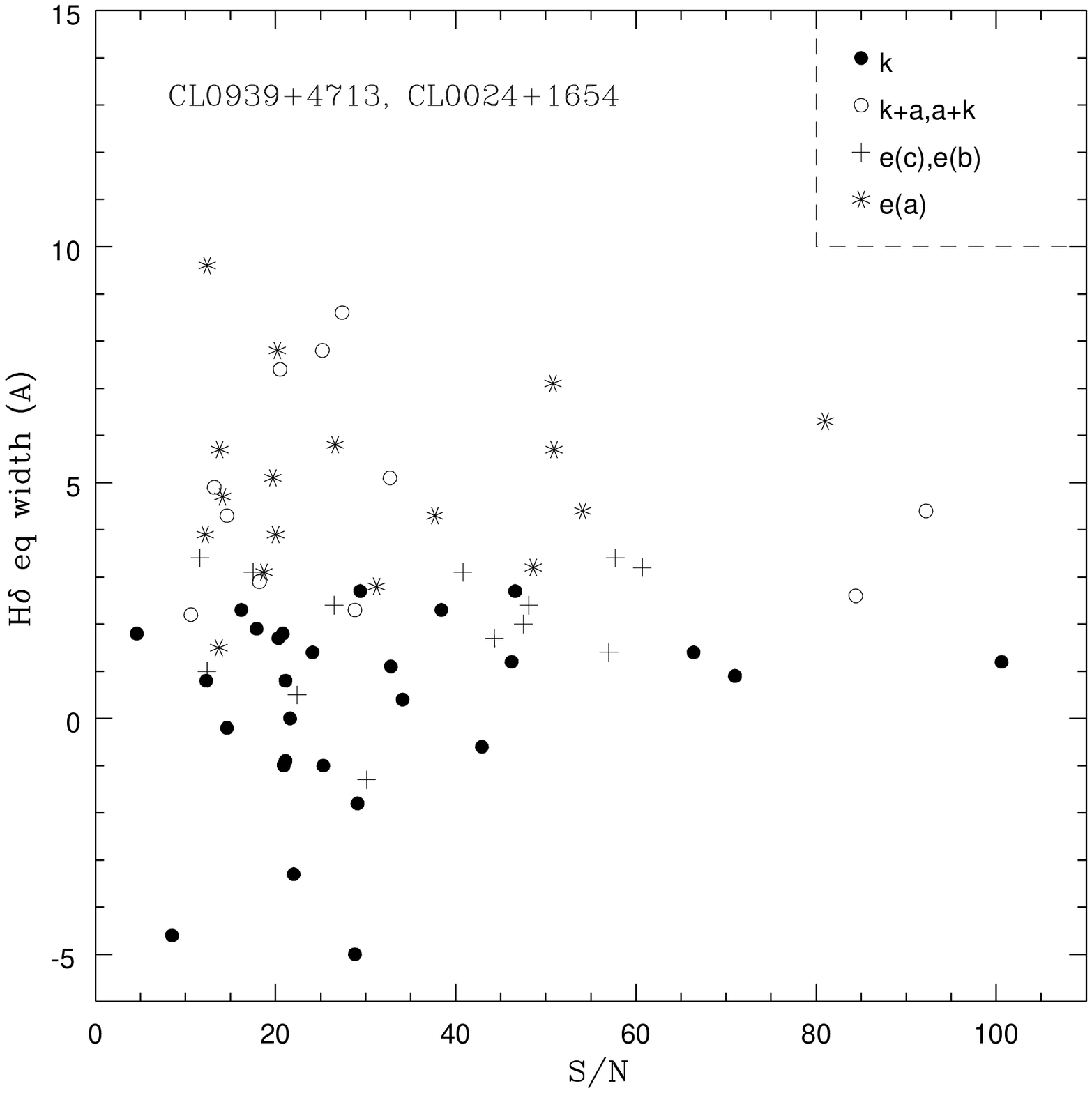,angle=0,width=3.0in}}
\noindent{\scriptsize
\addtolength{\baselineskip}{-3pt} 
\hspace*{0.3cm}

Fig.~1.\ a) Measured [O{\sc ii}] equivalent widths versus the S/N in
the straddling continuum.  The data are for 79 cluster members with
$q\leq 3$ in the clusters Cl\,0939$+$47 and Cl\,0024$+$16.  There is a
clean separation at $\sim -4$\AA\ between the types for which [O{\sc
ii}] was found by inspection and those for which it was judged to be
absent. b) The equivalent plot for the measured H$\delta$ equivalent
widths.  H$\delta$ strengths greater than 3\AA\ are clear detections,
according to the distribution, with 2--3\AA\ strengths ambiguous,
particularly at low S/N.

\addtolength{\baselineskip}{3pt}
}
\end{figure*}

The results of these tests are shown in Fig.~1a and Fig.~1b, where we
have plotted the equivalent widths as a function of signal-to-noise
ratio in the continuum bands straddling the feature. In Fig.~1b we show
that the galaxies that were designated by inspection as emission line
types all have [O{\sc ii}] EW stronger than $-3$\AA, while those
designated as having no emission lines (spectral types: k, k+a, or a+k,
see \S3.3) have [O{\sc ii}] EW weaker than $-4$\AA.  In fact, the
latter are consistent with non-detections: for 37 non-emission line
members, the median EW is +0.4\AA\ with quartiles of $-1.0$ to +2.6.
There is only a weak trend with signal-to-noise ratio. We conclude from
these data that we are complete for [O{\sc ii}] stronger than $-5$\AA,
with a high level of completeness down to $-3$\AA.  In other words,
even at the modest signal-to-noise ratios of these spectra, none of the
galaxies classified as non-emission types are likely to have emission
at greater than the $-3$\AA\ level, and certainly none have emission
stronger than $-5$\AA~(this limit corresponds to ``absent'' in Table~6).

In Fig.~1b we show a similar diagram for the same sample, this time for
H$\delta$.  Because it is weaker and in absorption, H$\delta$ is a more
difficult feature to measure; this is apparent from the stronger trend
with signal-to-noise ratio.  However, as for [O{\sc ii}], the
separation of those galaxies which are designated by inspection as
having moderate Balmer line strengths (k+a, a+k, and e(a), see \S 3.3)
from the non-Balmer galaxies (k and e(c) and e(b) types), is confirmed
by the objective measurements.  The boundary is around 2--3 \AA, below
which point we are unable, except at high S/N $>$ 50, to discern the
difference between the presence or absence of H$\delta$.  We conclude
from these data that we are complete above equivalent widths of +5\AA,
and mostly complete above +3\AA.  It is worth commenting that some of
the points with large negative equivalent widths for H$\delta$ arise
from strange continuum levels, rather than from the feature seen in
emission (although there is at least one clear case of H$\delta$ in
emission, a rare phenomena among luminous galaxies).

\subsection{Cluster Membership}

As was noted above, Table~4 is split into two parts on the basis of
whether a galaxy is classed as a ``Cluster'' member or ``Field''.  To
accomplish this we define redshift ranges for the various clusters;
these ranges are purposefully chosen to be large to ensure that we
retain any galaxies in the large-scale  structure surrounding the
clusters, while at the same time minimizing  the contamination by field
galaxies.  In Fig.~2 we show the redshift distributions for the
individual cluster fields; in each panel the inset provides a more
detailed view of the velocity distribution close to the cluster mean.
The bin size in these plots has been arbitrarily chosen and may
artificially enhance or suppress the visibility of any structures
within the clusters.  We list the resulting mean redshift, restframe
velocity dispersion and redshift range defining each cluster in
Table~2.  We reiterate that the velocity dispersions are likely to be
overestimates of the true dispersion of the well-mixed cluster
population.   We also list in Table~2 the number of member galaxies in
our catalog for each cluster.  Using these definitions our catalog
contains a total of 424 cluster members and 233 field galaxies.

%
%
\begin{figure*}
\centerline{\psfig{file=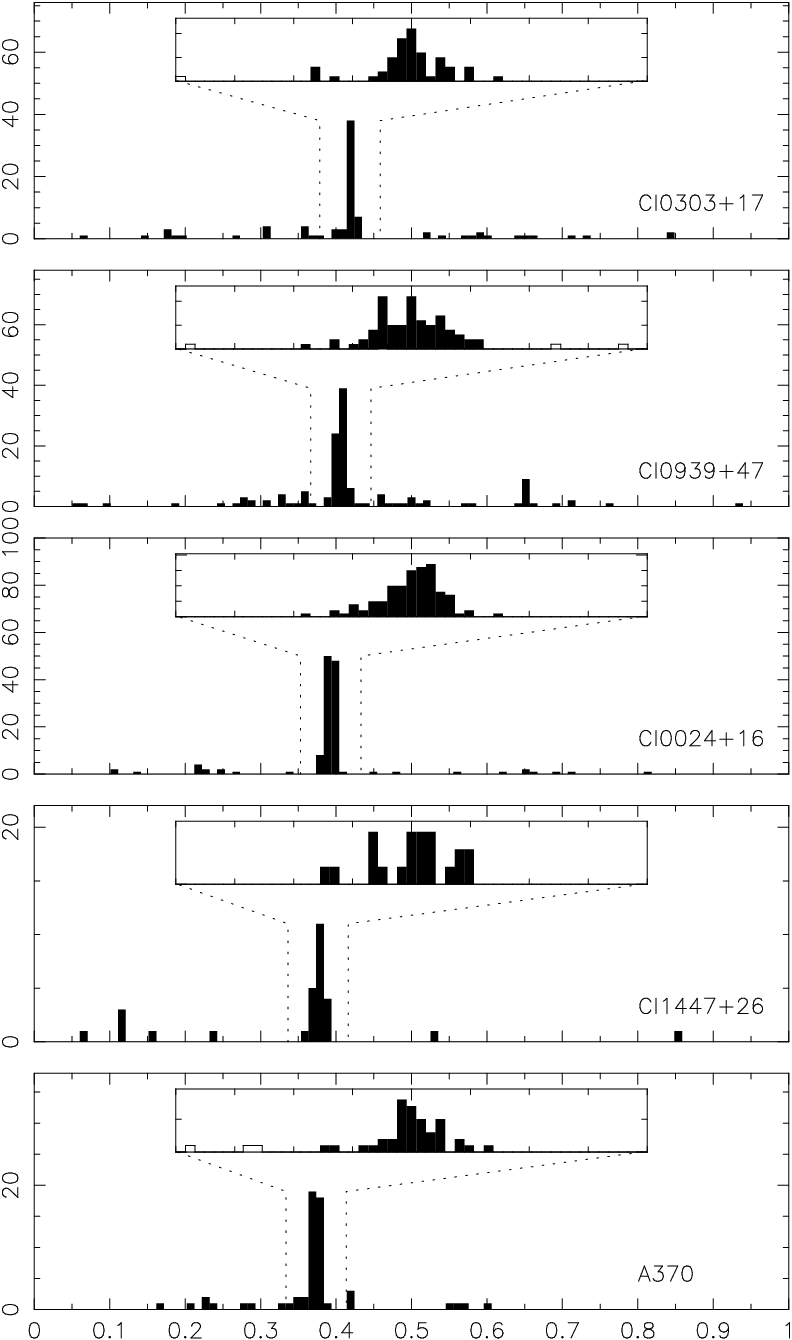,angle=0,width=2.5in}\hspace{0.5in}
\psfig{file=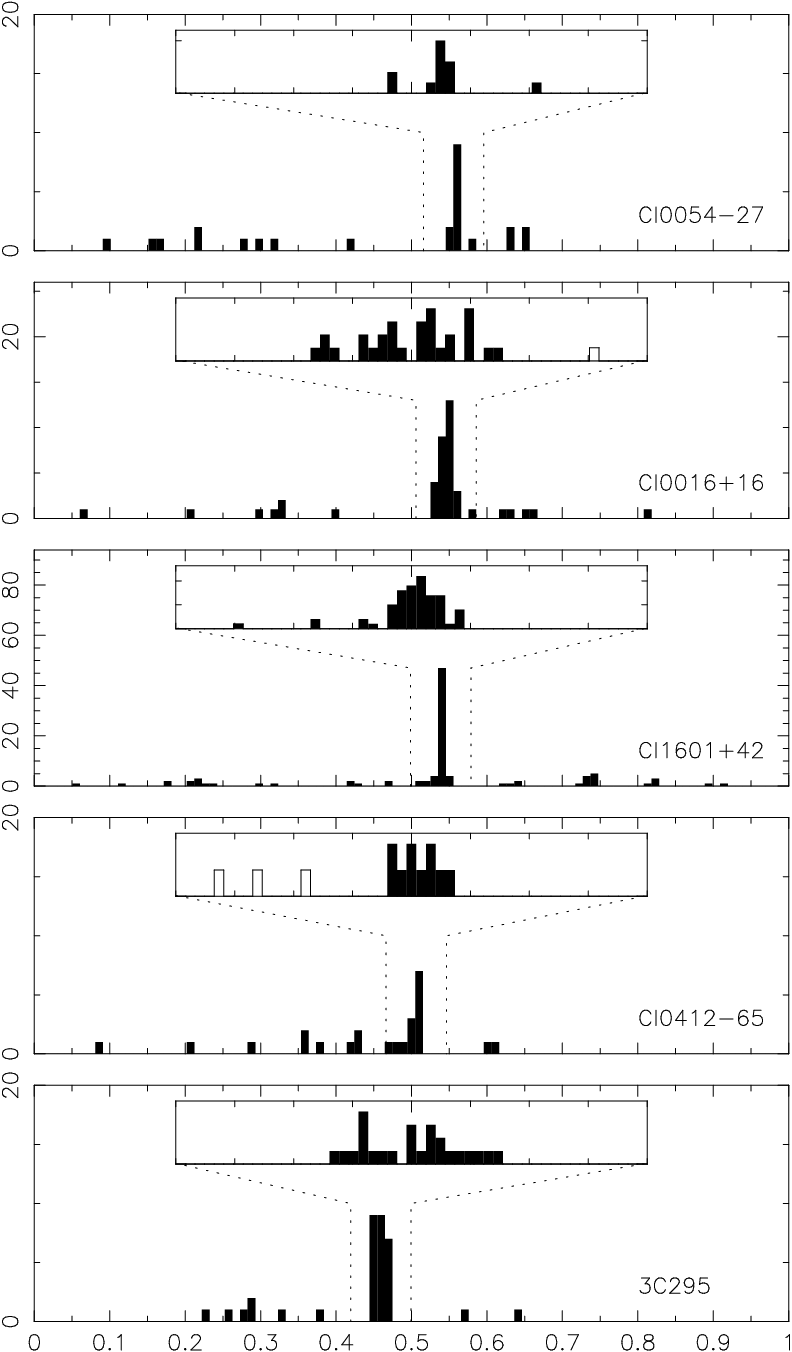,angle=0,width=2.5in}}

\noindent{\scriptsize
\addtolength{\baselineskip}{-3pt} 

\hspace*{0.3cm} 
Fig.~2.\ Redshift distributions for the fields of our
10 clusters.  We show the redshift range $z=0$--1 in the full plot and
then, in the insets, an expanded region (width $\Delta z = 0.08$)
centered on the cluster redshift, in the inset panels we show the
cluster members as a filled histogram and the field galaxies as open.
The redshift axis in the inset panel is marked with $\Delta z=0.01$
increments and the vertical axis is the same as the main panel.

\addtolength{\baselineskip}{3pt}
}
\end{figure*}

The redshift distribution for all galaxies classed as field is shown as
the open histogram in Fig.~3; the galaxies with {\it HST} morphologies
are shown as the filled histogram. The median redshift of the whole
field sample is $<\! z\! >=0.42$, while for the morphological
sub-sample it is slightly higher at $<\! z\! >=0.46$ (Fig.~3). These
values are very similar to the median redshift of our 10 clusters,
$<\!  z\!  >=0.44$, allowing us to easily compare the broad properties
of the cluster and field samples. A total of 20 stars were observed
(all in either the flanking fields or from the earlier DG92
observations); these are included at the bottom of Table~4b, but we do
not discuss them further.

In Fig.~3 we may be seeing some evidence for a deficit in the total
field redshift distribution, between $z\sim 0.4$--0.6, which would
result from the inclusion of a few field galaxies in the cluster
catalog. This would include galaxies in the supercluster environment,
if any, in which the clusters reside, or truly unassociated galaxies
relatively far from the cluster but within the wide velocity limits
imposed by the cluster's velocity dispersion.  To estimate the extent
of this effect we use two approaches.  Firstly, a conservative upper
limit on the deficit  comes from linearally extrapolating the trends of
number versus redshift in the field  at $z<0.35$ and $z>0.60$ to limit
the likely number of field galaxies in the intervening redshift range.
~From this we estimate that there should be $\ls 160$ field galaxies in
the range $z=0.3$--0.6, compared to the observed number of 92, giving
an upper limit on the deficit of $\sim 70$ galaxies, or $\ls 7$ per
cluster.  Alternatively, using the regions where the redshift limits of
the cluster and field samples overlap between different clusters we
estimate the contamination from random, unrelated field galaxies
is of the order of $1.0\pm 0.7$ galaxies per cluster in our largest
velocity range.  We conclude therefore that the contamination from
galaxies unrelated to the cluster, or its supercluster, does not exceed
7 galaxies per cluster and is probably closer to 1--2 galaxies.

\subsection{Spectral Classification}

To assess the distribution in the star-formation properties of galaxies
in our catalog we have found it useful to classify their spectra into a
number of classes.  These classes are broadly based upon those used by
DG92 and CS87, however, the number of classes has been expanded to
better cover the full range of features seen in our large sample.  We
have also used the properties of low redshift integrated spectra
(Kennicutt 1992) and the expected characteristics from spectral
modelling to help us define the limits of some of the classes.  In
revising the classifications we therefore found it necessary to
redefine some of the boundaries previously used for the spectral
classes.  Hence, to reduce confusion between our new classes and those
used previously we adopt a new nomenclature and give this and the
details of the classification scheme in Table~6.  We show a schematic
representation of this spectral classification in Fig.~4.  It should be
noted that for those spectra where sky residuals or the available
spectral range precluded the observation of one of the diagnostic
spectral features, we have made used the strength of the other Balmer
series lines (if H$\delta$ was unobservable) or emission lines (if
[O{\sc ii}] was unobservable) to identify the most likely spectral
class.  In the few cases where this has been done comments are included
in Table~4.

%
%
\hbox{~}
\centerline{\psfig{file=f3.ps,angle=0,width=2.5in}}
\noindent{\scriptsize
\addtolength{\baselineskip}{-3pt} 
\hspace*{0.3cm}Fig.~3.\ Redshift distribution for galaxies classed as
non-members in the fields of the 10 clusters. The open histogram gives
the total redshift distribution for the field galaxies (233 galaxies),
the filled histogram is those field galaxies which lie within the {\it WFPC2}
field and for which we therefore have detailed morphological
information (71 galaxies).

\addtolength{\baselineskip}{3pt}
}

%
%
\hbox{~}
\centerline{\psfig{file=f4.ps,angle=0,width=2.5in}}
\noindent{\scriptsize
\addtolength{\baselineskip}{-3pt} 
\hspace*{0.3cm} Fig.~4.\ A schematic representation of the spectral
classification scheme used in this work.  We show the regions of the
H$\delta$--[O{\sc ii}] equivalent width plane populated by the various
spectral types.  Those spectral classes not based upon the line
strengths of H$\delta$ and [O{\sc ii}] (e.g.\ CSB, e(n), etc.) are not
marked. 

\addtolength{\baselineskip}{3pt}
}

Briefly the overlap between the new system and previous ones can be
summarized as follows: we retain the general features of the DG92
system, including k-type and the general class of ``e'' (emission)
galaxies.  However, we replace the mixed nomenclature ``E+A'' with k+a
(following the suggestion of Franx 1993) and a+k, depending on the
strength of the H$\delta$ Balmer line.  We also subdivide emission line
galaxies into e(a) types (with strong Balmer absorption), e(c) for
those with weak or moderate Balmer absorption, and e(b) for those with
very strong [O{\sc ii}] (this can sometimes be combined with e(a) for
galaxies with both strong [O{\sc ii}] emission and strong Balmer
absorption). This nomenclature reflects the nature of the spectra, with
e(a) indicating a population of A stars, e(b) a spectrum similar to
that expected for a burst of star-formation and e(c) a spectrum for a
system undergoing a more constant SFR.

%
%
\begin{table*}
{\scriptsize
\begin{center}
\centerline{\sc Table 6}
\vspace{0.1cm}
\centerline{\sc Spectral Classification Scheme}
\vspace{0.3cm}
\begin{tabular}{lcccl}
\hline\hline
\noalign{\smallskip}
 {Class} & {EW [O{\sc ii}]\,3727} & {EW H$\delta$} &  Color & Comments \cr
 & (\AA) & (\AA) & &  \cr
\hline
\noalign{\smallskip}
k & absent & $< 3$ & ... & passive \cr
k+a & absent & 3--8 & ... & moderate Balmer absorption without emission \cr
a+k & absent & $\geq 8$ & ... & strong Balmer absorption without emission \cr
e(c) & yes,$< 40$ & $< 4$ & ... & moderate Balmer absorption plus emission, spiral-like \cr
e(a) & yes & $\geq 4$ & ... & strong Balmer absoprtion plus emission \cr
e(b) & $\geq 40$ & ... & ... & starburst \cr
e(n) & ... & ... & ... & AGN from broad lines or [O{\sc iii}]\,5007/H$\beta$ ratio \cr
e &  yes & ? & ... & with at least one emission line but S/N too low to classify \cr
? &   ?  & ? & ... & unclassifiable \cr
\noalign{\smallskip}
CSB & ... & ... & very blue & photometrically-defined starburst \cr 
\noalign{\smallskip}
\noalign{\hrule}
\noalign{\smallskip}
\end{tabular}
\end{center}
}
\vspace*{-0.8cm}
\end{table*}

In comparison to other earlier work, the PSG and the HDS galaxies of
CS87 fall mostly into the a+k and k+a classes. The CS87 ``Spiral''
types are placed in e(c) and e(a); however, the SB galaxies are not the
same as our type e(b), because the criteria for these in CS87 was not
based on [O{\sc ii}] strength.  We note that spectral classes described
in Table~6 can be grouped in three main categories:  passive (k); past
star-forming (k+a and a+k) and currently star-forming (e(c), e(a) and
e(b)). AGN spectra (e(n)) are excluded in this division (Table~6).

In Column~8 of Table~4 we include a photometric classification in the
case of the bluest galaxies.  These are labeled ``Color Starburst''
(CSB) if their restframe color is bluer than that expected for a low
metallicity model galaxy with an increasing star-formation rate (P98).
This allows us to conservatively identify those galaxies whose very
blue colors can only be explained with a current starburst, whatever
their spectral type may be.

%
%
\hbox{~}
\centerline{\psfig{file=f5.ps,angle=0,width=2.5in}}
\noindent{\scriptsize
\addtolength{\baselineskip}{-3pt} 
\hspace*{0.3cm}
Fig.~5.\ Representative spectra from each of the spectral classes in
our adopted scheme (Table~6, Fig.~4).  These are plotted with arbitrary
vertical scaling and in the restframe.  The galaxies are all cluster
members with $q=1$ and come from Cl\,0939$+$47 and Cl\,0024$+$16.  The
spectra are not fluxed.

\addtolength{\baselineskip}{3pt}
}

To better illustrate the properties of the new classification scheme we
show in Fig.~5 a high-quality, representative member of each class from
our catalog.  In Table~7 we give the distribution of spectral classes
within the different clusters (for $q\leq 4$), as well as the total
numbers across all the clusters and the equivalent values for our field
samples. As can be seen, the clusters are populated by a wide variety of
spectral classes, although comparisons between clusters are not simple
owing to the different apparent magnitudes of the samples and the
attending variation in the typical quality of the spectra.  Table~7
also lists the equivalent numbers of galaxies in each spectral
class for which we have morphological information.

%
%
\begin{table*}
{\scriptsize
\begin{center}
\centerline{\sc Table 7}
\vspace{0.1cm}
\centerline{\sc Spectral Samples}
\vspace{0.3cm}
\begin{tabular}{lcccccccccc}
\hline\hline
\noalign{\smallskip}
 {Cluster} & N$_{\rm tot}$ & k & k+a & a+k & e(a) & e(c) & e(b) & e(n) & e & ? \cr
\hline
\noalign{\smallskip}
\multispan{11}{\hfil{Full~Sample}\hfil}\cr
\noalign{\smallskip}
 Field         &  233 &   ~36 &   ~7 &   ~0 &   37 &   74 &   39 &   ~3 &   25 &   12 \cr
 \noalign{\smallskip}
 A\,370        &  ~40 &   ~26 &   ~1 &   ~0 &   ~3 &   ~8 &   ~1 &   ~0 &   ~0 &   ~1 \cr
 Cl\,1447$+$26 &  ~21 &   ~~7 &   ~1 &   ~0 &   ~6 &   ~6 &   ~1 &   ~0 &   ~0 &   ~0 \cr
 Cl\,0024$+$16 &  107 &   ~47 &   12 &   ~2 &   13 &   21 &   ~6 &   ~2 &   ~2 &   ~2 \cr
 Cl\,0939$+$47 &  ~71 &   ~31 &   13 &   ~6 &   ~7 &   10 &   ~2 &   ~1 &   ~1 &   ~0 \cr
 Cl\,0303$+$17 &  ~51 &   ~14 &   ~4 &   ~4 &   ~6 &   12 &   ~6 &   ~0 &   ~5 &   ~0 \cr
 3C\,295       &  ~25 &   ~10 &   ~6 &   ~1 &   ~2 &   ~2 &   ~0 &   ~3 &   ~1 &   ~0 \cr
 Cl\,0412$-$65 &  ~10 &   ~~0 &   ~1 &   ~0 &   ~0 &   ~0 &   ~1 &   ~0 &   ~1 &   ~7 \cr
 Cl\,1601$+$42 &  ~58 &   ~33 &   15 &   ~0 &   ~4 &   ~3 &   ~2 &   ~1 &   ~0 &   ~0 \cr
 Cl\,0016$+$16 &  ~29 &   ~13 &   ~6 &   ~5 &   ~3 &   ~1 &   ~1 &   ~0 &   ~0 &   ~0 \cr
 Cl\,0054$-$27 &  ~12 &   ~~5 &   ~1 &   ~0 &   ~0 &   ~1 &   ~0 &   ~0 &   ~3 &   ~2 \cr
 \noalign{\smallskip}
 Total        &   424 &   186 &   60 &   18 &   44 &   64 &   20 &   ~7 &   13 &   12 \cr
\noalign{\medskip}
\multispan{11}{\hfil{Morphological~Sample}\hfil}\cr
\noalign{\smallskip}
 Field         &   ~71 &   ~11 &   ~5 &   ~0 &   ~7 &   25 &   11 &   ~0 &  ~9 &  ~3 \cr
 \noalign{\smallskip}
 A\,370        &   ~14 &   ~~9 &   ~0 &   ~0 &   ~2 &   ~1 &   ~1 &   ~0 &  ~0 &  ~1 \cr
 Cl\,1447$+$26 &   ~10 &   ~~4 &   ~0 &   ~0 &   ~2 &   ~3 &   ~1 &   ~0 &  ~0 &  ~0 \cr
 Cl\,0024$+$16 &   ~42 &   ~25 &   ~6 &   ~0 &   ~4 &   ~3 &   ~3 &   ~0 &  ~1 &  ~0 \cr
 Cl\,0939$+$47 &   ~31 &   ~16 &   ~6 &   ~2 &   ~5 &   ~2 &   ~0 &   ~0 &  ~0 &  ~0 \cr
 Cl\,0303$+$17 &   ~28 &   ~~9 &   ~2 &   ~3 &   ~4 &   ~5 &   ~4 &   ~0 &  ~1 &  ~0 \cr
 3C\,295       &   ~20 &   ~~7 &   ~5 &   ~0 &   ~2 &   ~2 &   ~0 &   ~3 &  ~1 &  ~0 \cr
 Cl\,0412$-$65 &   ~~2 &   ~~0 &   ~0 &   ~0 &   ~0 &   ~0 &   ~0 &   ~0 &  ~0 &  ~2 \cr
 Cl\,1601$+$42 &   ~28 &   ~20 &   ~5 &   ~0 &   ~0 &   ~1 &   ~1 &   ~1 &  ~0 &  ~0 \cr
 Cl\,0016$+$16 &   ~22 &   ~10 &   ~4 &   ~5 &   ~2 &   ~1 &   ~0 &   ~0 &  ~0 &  ~0 \cr
 Cl\,0054$-$27 &   ~~7 &   ~~4 &   ~0 &   ~0 &   ~0 &   ~0 &   ~0 &   ~0 &  ~1 &  ~2 \cr
 \noalign{\smallskip}
 Total         &   204 &   104 &   28 &   10 &   21 &   18 &   10 &   ~4 &  ~4 &  ~5 \cr
\noalign{\smallskip}
\noalign{\hrule}
\noalign{\smallskip}
\end{tabular}
\end{center}
}
\vspace*{-0.8cm}
\end{table*}

\section{Basic Properties and Correlations of the Data}

To start the discussion of the spectroscopic sample we have assembled,
we review the basic properties of the sample as a whole.  We focus on
a few of the correlations between the various properties of the galaxies in
the sample, in particular the relationships between the morphological,
spectral, and kinematic characteristics of certain classes of cluster
galaxies.  In the following discussion we will include the uncertain
spectral classes (marked with a ``:'' in Tables~4), unless otherwise
stated.

\subsection{Luminosity Functions for the Morphological Classes}

In order to draw conclusions from our spectroscopic study in the
context of the broader morphological catalog (S97), we need to compare the
sampling in absolute magnitude of the two catalogs.  Fig.~6a shows the
absolute magnitude distribution for galaxies in the spectroscopic
catalog for which ground-based $r$-band photometry is available. This
filter approximates $V$ in the restframe for all 10 clusters.  Our
assumption of a single {\it K}-correction (from an spectral energy
distribution (SED) corresponding to a present day Sbc) introduces only
small errors into the magnitude distribution ($\ls 0.06$ mags for E/S0
and Sd/Irr SED).  Fitting a Schechter function to the bright-end of the
distribution in Fig.~6a, we obtain a characteristic magnitude of
$M^\ast_V = (-20.64\pm 0.16) + 5 \log_{10} h$ (for a fixed faint-end
slope of $\alpha=-1.25$ as adopted in S97).  This is to be compared to a fit obtained to
the morphological counts in the cluster fields corrected for likely
contamination in the manner described in S97.  Fitting to the composite
luminosity function of all morphological types across the 10 clusters we find  
$M^\ast_V = (-20.79\pm 0.02) + 5 \log_{10} h$ (for $\alpha = -1.25$).
This good agreement indicates that the spectroscopic catalog fairly
samples the morphological catalog for $M_V < -19 + 5 \log_{10} h$.
 
Fig.~6b shows how the spectroscopic sampling compares as a function
of morphological type within the clusters.
This is achieved by comparing the spectroscopic
sample for $M_V < -19 + 5 \log_{10} h$ to the field-corrected morphological
counts of S97.  There is no significant trend with morphological type
except for the selection effect, discussed in \S2.1, built into the
original sample selection:  the Sd/Irregular galaxies are oversampled
relative to the E--Sc types (although there is considerable uncertainty
in the statistical correction for field galaxies in this bin, S97).
This plot allows us to quantify and correct for the sample selection in
our analysis as required.

\subsection{Luminosity Functions for the Spectral Classes}

The absolute magnitude distribution of the spectral classes defined in
this paper will be important to understanding their relationships within the
framework of galaxy evolution models.  Fig.~7a shows that the magnitude
distributions brighter than $M_V = -19 + 5 \log_{10} h$ for
spectral classes k, k+a and a+k are statistically indistinguishable.
In contrast, the e(a) and e(c) classes appear to systematically fainter
than the k class; this difference is confirmed at the $\gs 95$\%
confidence limit using two-sample Kolmogorov-Smirnoff tests.  It is
important to keep in mind the ``completeness'' limit of the
spectroscopic catalog estimated in \S4.1, which means that these
differences could be larger, and, for example, the apparent peak in the
luminosity distribution of e(a)'s in Fig.~7a may be partly an artifact
of incomplete sampling.  The difference between the k class and 
the fainter e(b)
class is clearly significant: the likelihood that the two samples
are drawn from the same luminosity distribution is only $\log_{10} P
\sim -4.6$.  Again, the difference may be larger still, owing to the
incomplete sampling below ($M_V = -19 + 5 \log_{10} h$).  We know
of no selection effect in our study that would cause us to miss bright
e(b) cluster galaxies. As we discuss in P98, the fact that the
galaxies that we have identified as bursting are fainter than the other
classes is significant, and discouraging for models that attempt to
interpret these starbursts as progenitors for galaxies with strong
Balmer lines in their spectra.

%
%
\begin{figure*}
\centerline{\psfig{file=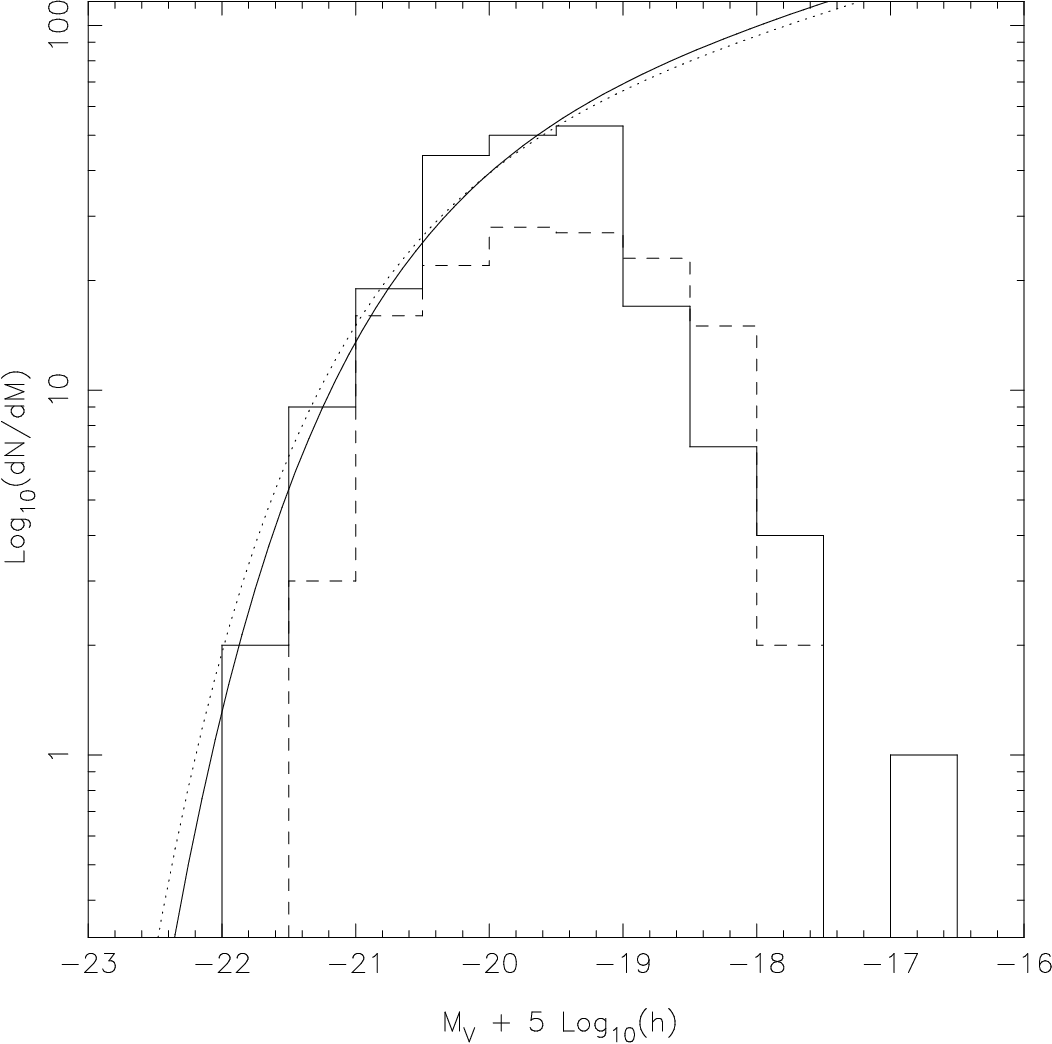,angle=0,width=2.5in}
\hspace*{0.5in}
\psfig{file=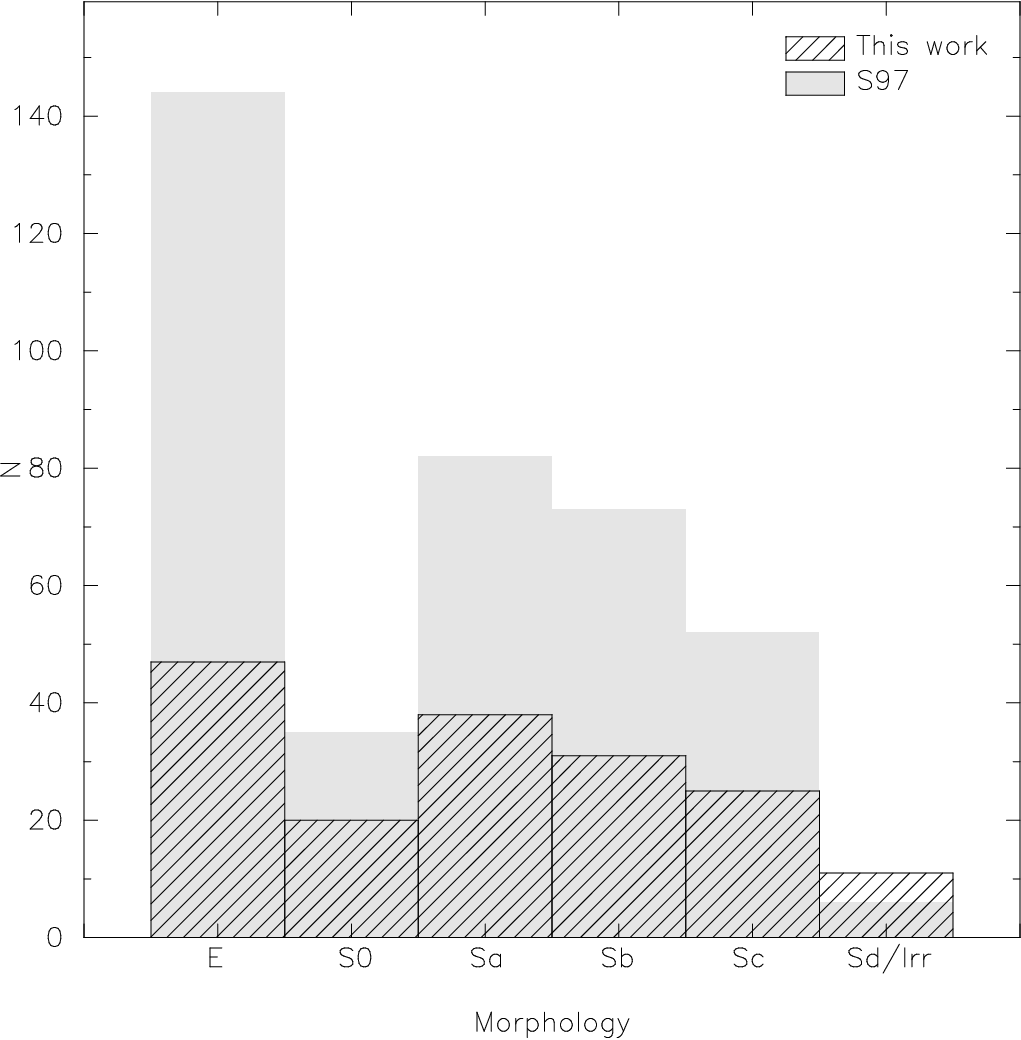,angle=0,width=2.5in}}

\noindent{\scriptsize
\addtolength{\baselineskip}{-3pt} 
\hspace*{0.3cm}
Fig.~6.\ a) The absolute magnitude distribution, in the restframe
$V$-band, for the spectroscopic sample of cluster (solid histogram) and
field (dashed histogram) galaxies (for those galaxies with DG92
photometry).  The solid curve is the best-fit Schechter function to the
cluster members, using a fixed faint-end slope of $\alpha=-1.25$.  The
characteristic luminosity derived from the fit is $M^\ast_V = -20.64\pm
0.16 + 5 \log_{10} h$.  The dotted line shows the fit to the luminosity
functions derived from the morphological counts in the frames,
corrected for field contamination (S97).  This fit is shown for
$\alpha=-1.25$ and with arbitrary vertical scaling.  The good agreement
of the two distributions shows that the spectroscopic catalog provides a
representative luminosity distribution in the clusters at $M_V < -19 +
5 \log_{10} h$.  b)  The numbers of the different morphological types
in the spectroscopic catalog (hatched histogram) brighter than $M_V =
-19 + 5 \log_{10} h$.  The filled histogram indicates the total numbers
expected from the observed morphological counts  in the clusters to the
same depth, after correcting for field contamination (see S97).  

\addtolength{\baselineskip}{3pt}
}
\end{figure*}

The cluster sample defined by our redshift measurements also allow us
to unambiguously derive, for the first time, the absolute magnitude
distributions as a function of morphology, again for $M_V \leq -19
+ 5 \log_{10} h$. Fig.~7b shows a broad similarity between the absolute
magnitudes of early- and mid-type disk systems (S0--Sa--Sb--Sc).
Compared to these, elliptical galaxies show a systematically brighter
distribution, and irregular galaxies exhibit a tail of fainter
systems.  These trends are in good agreement with what is seen in
low-redshift clusters.

%
%
\begin{figure*}
\centerline{\psfig{file=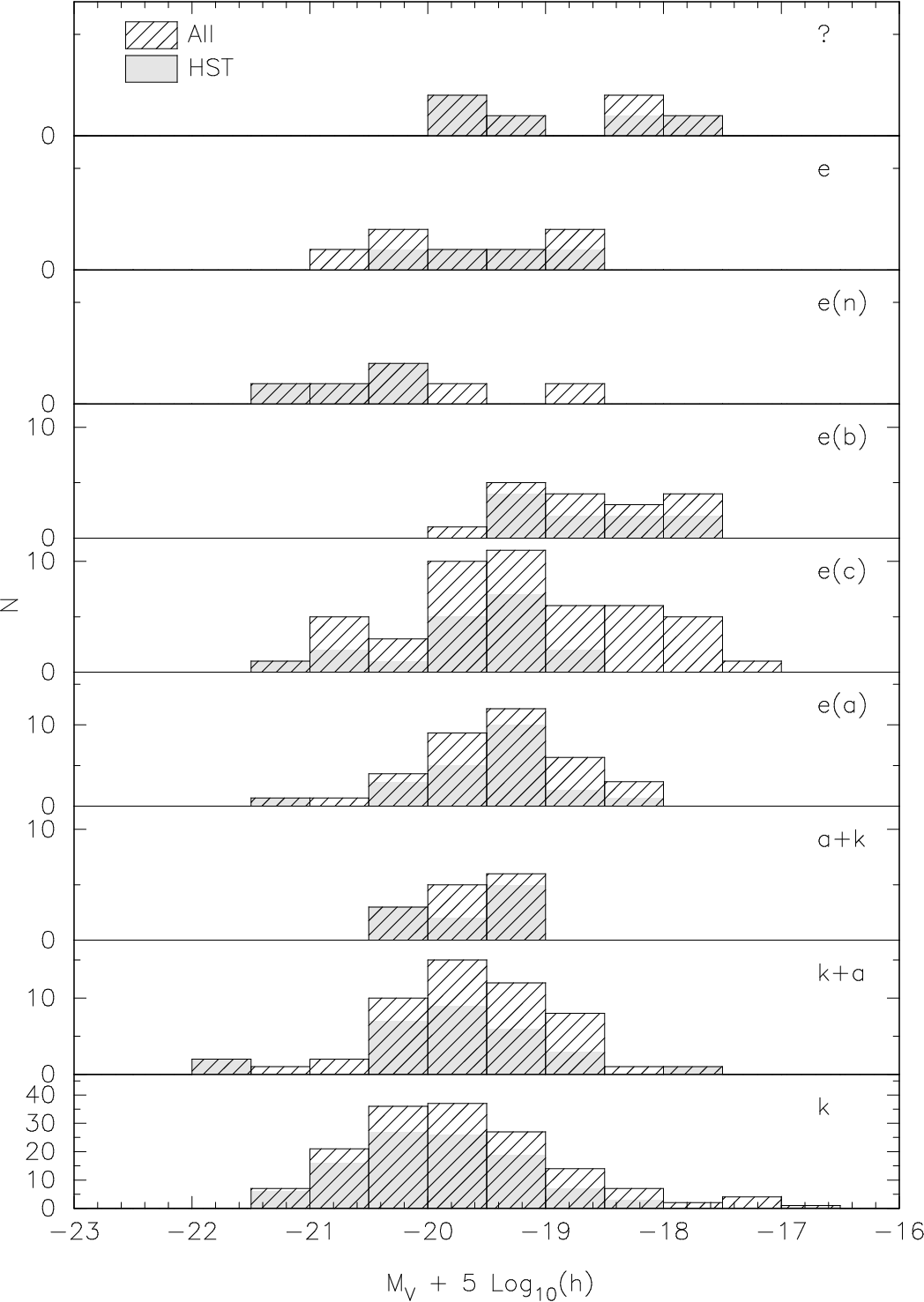,angle=0,width=2.5in}
\hspace*{0.5in}\psfig{file=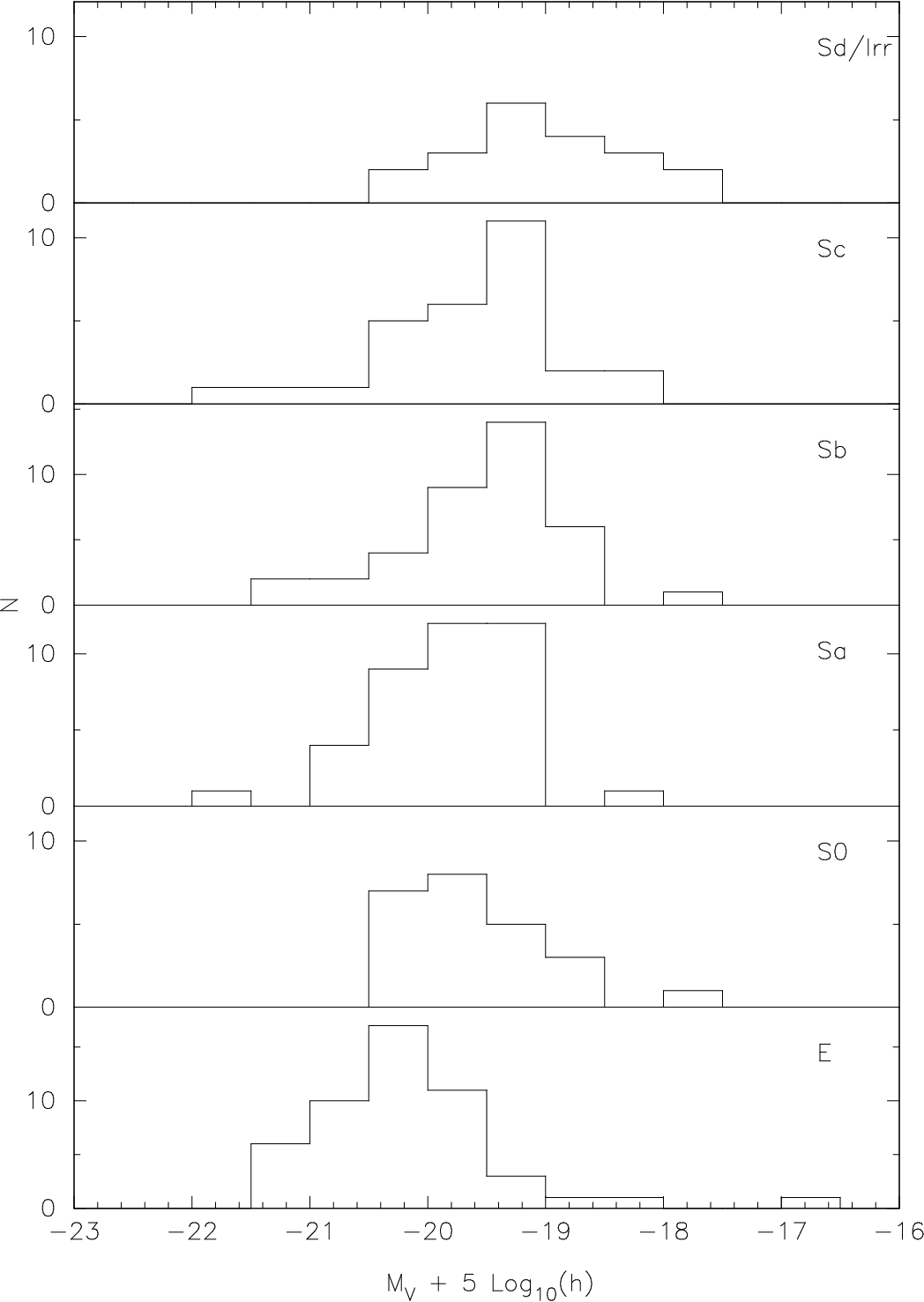,angle=0,width=2.5in}}
\noindent{\scriptsize \addtolength{\baselineskip}{-3pt} \hspace*{0.3cm}
Fig.~7.\ a) The absolute magnitude distribution in the rest-frame
$V$-band for cluster members separated into  the different
spectroscopic classes for both the {\it HST} fields and the full
sample.  In this panel we convert the ground-based $r$-band photometry
to rest-frame $V$-band.  b) The same distribution, but now separated on
morphological type and using the {\it HST} photometry.  The {\it
K}-correction applied in both cases assumes a spectral energy
distribution (SED) similar to a local Sbc galaxy, this introduce a
typical systematic error of only $\sim 0.06$ mags for E/S0 and Sd/Irr SEDs.

\addtolength{\baselineskip}{3pt}
}
\end{figure*}

%
%
\begin{figure*}
\centerline{\psfig{file=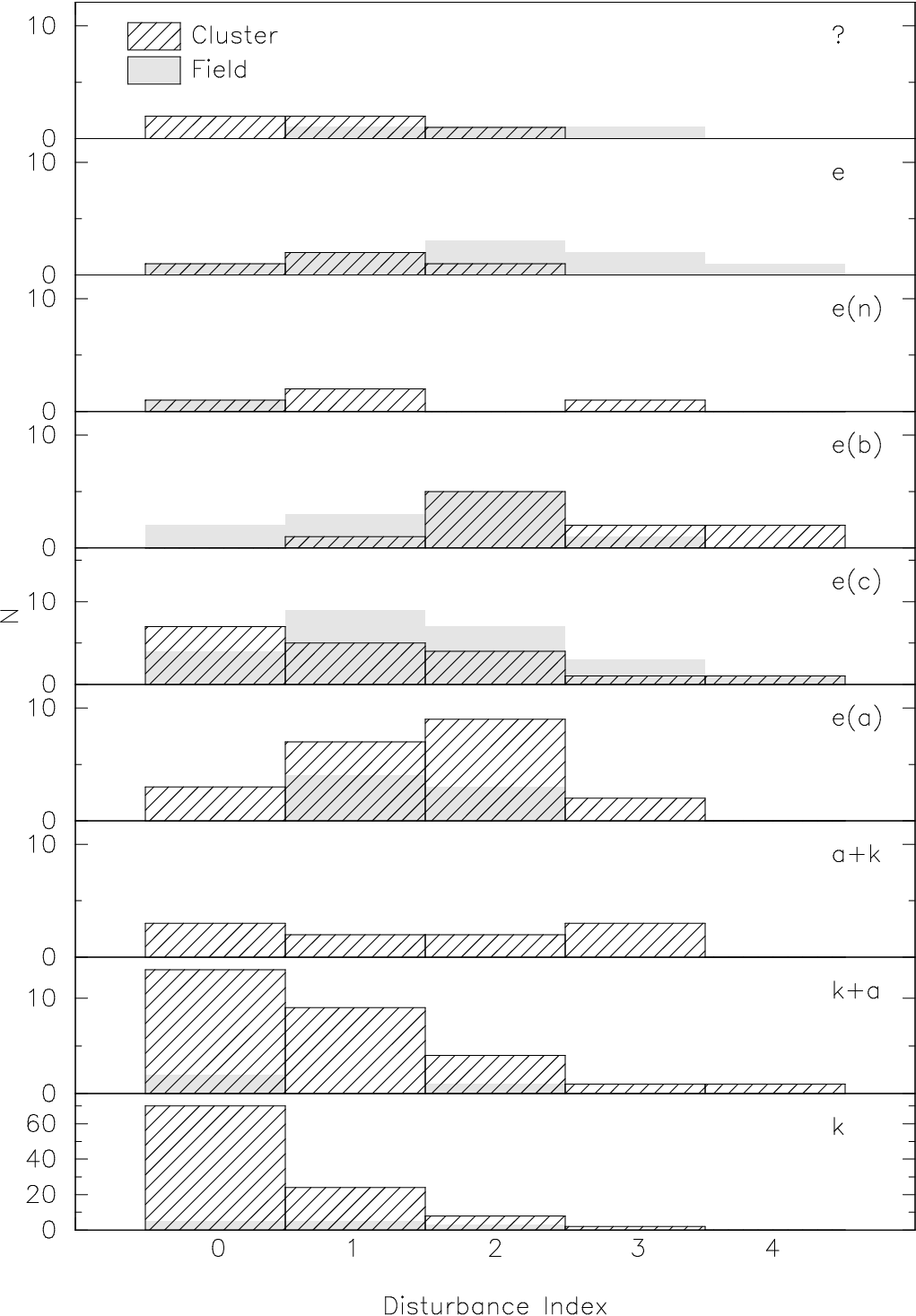,angle=0,width=2.in}
\hspace*{0.5in}
\psfig{file=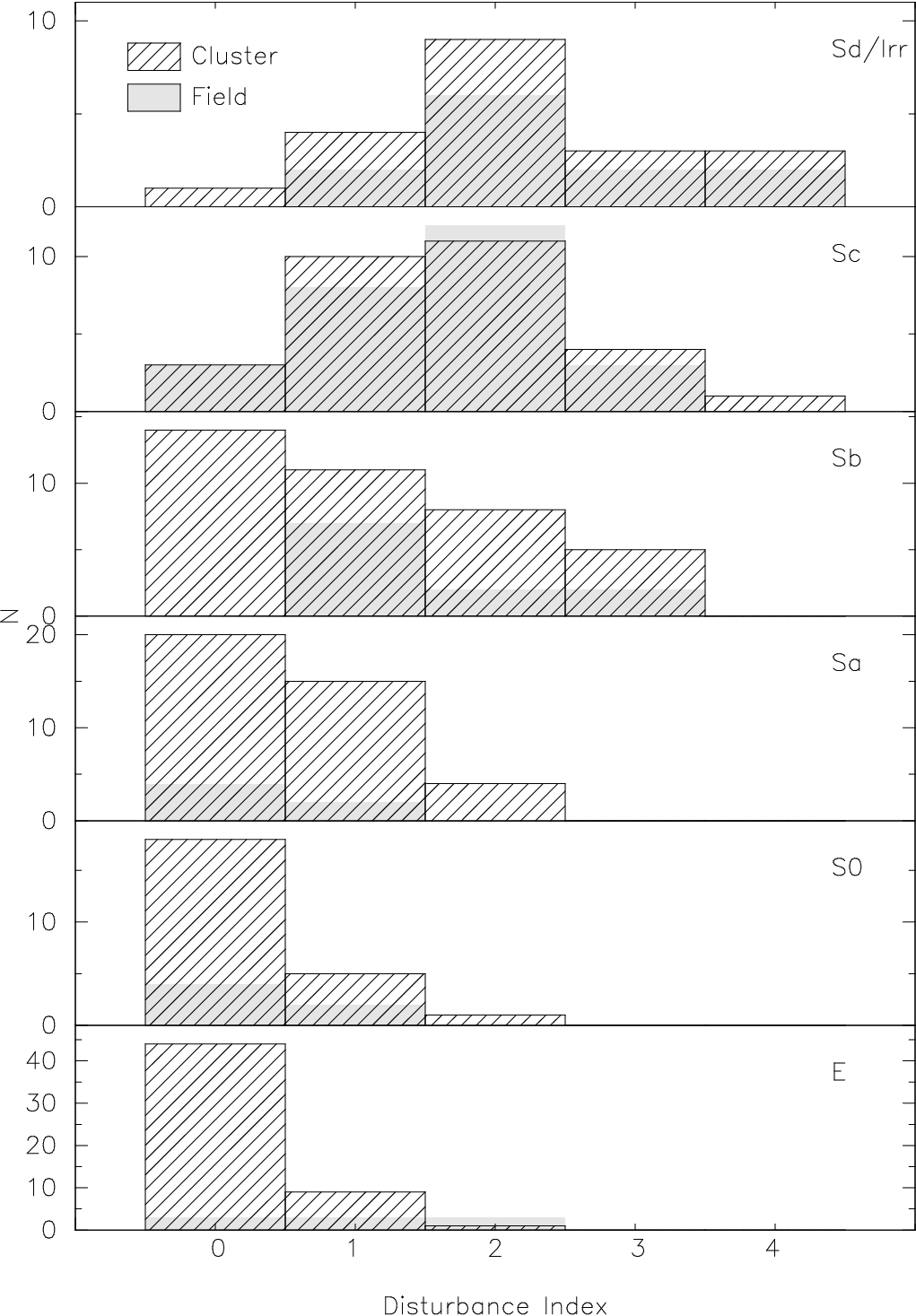,angle=0,width=2.in}
}

\noindent{\scriptsize
\addtolength{\baselineskip}{-3pt} 
\hspace*{0.3cm}
Fig.~9.\ a) Histograms indicating the distribution of spectral types
within the different disturbance classes, $D$, for both cluster and
field populations (hatched and filled histograms respectively). $D\geq
2$ denotes strongly a asymmetric or disturb light distribution.  b)
The numbers of galaxies in the different disturbance classes as a
function of galaxy morphology.  Note the broad similarity of the
cluster and field distributions for the later-type spiral galaxies.

\addtolength{\baselineskip}{3pt}
}
\end{figure*}

\subsection{Morphological Properties of the Cluster Galaxies}

What do the galaxies in our spectral classes look like?  We illustrate
the morphologies of the cluster members within each spectral class in
Figs.~8.  The general trend towards later-types in the active spectral
classes is clear.  The passive spectral classes are dominated by
early-type galaxies, particularly ellipticals.  The correspondence of
morphology and spectral properties, the same as found for low-redshift
analogs. indicates that a substantial fraction of the luminous
ellipticals of these clusters was in place by $z\sim 0.5$ (Ellis et
al.\ 1997, Dressler et al.\ 1997).  The e(c) spectra are generally
associated with disk galaxies, most of them familiar spirals and
irregulars.  This is true of some of the e(a)'s as well, but this class
also includes many disk systems that look more disturbed than typical
present-day spirals.  The k+a/a+k class does include some elliptical
galaxies, but the majority are disk galaxies, a few of which
have an irregular or disturbed appearance. The significance of the
correlation of morphology and spectral class are discussed further
in B98.

%
%
\centerline{\psfig{file=f10.ps,angle=0,width=2.5in}
}

\noindent{\scriptsize
\addtolength{\baselineskip}{-3pt} 
\hspace*{0.3cm}
Fig.~10.\ A comparison of the distribution of morphological type
within each spectroscopic class, for both cluster and field galaxies.

\addtolength{\baselineskip}{3pt}
}

We briefly discuss evidence for interactions and mergers on
the spectral classes of galaxies in our cluster samples.  (We also
comment on this issue in \S5.2 which deals with the kinematics of the
different cluster populations.)   We show in Fig.~9a the distribution
of disturbance class within the different spectral classes.  The image
disturbance, $D$, is a visual classification of the degree to which
the galaxy's structure appears distorted or disturbed (S97) {\it
compared to a typical low-redshift galaxy of the same morphological
type}.  The $D$ class correlates well with the asymmetry of the
galaxy's light profile (S97).   Fig.~9a suggests that the spectral
properties of the galaxies broadly correlate with the degree of image
distortion and disturbance, the active and recently active populations
having more galaxies classed as strongly asymmetric or distorted.
However, looking at Fig.~9b we see an arguably stronger correlation
between morphology and $D$ with a pronounced shift towards higher $D$
values in going to later-types (Sb--Sd/Irr).  This could be due to a
failure on our part to actually separate disturbance from a natural
trend toward more irregular morphology for late-type systems, but the
large number of $D\geq 2$ Sc galaxies (a type that is generally
symmetric for low-redshift galaxies) suggests that the effect is real.\footnote{This tendancy of intermediate-redshift disk galaxies to appear
more asymetric than low-redshift galaxies of similar type has been
reported in essentially all studies of this type}.
If so, it most likely reflects the greater fragility of disks
(compared to bulges) to perturbations, and the greater frequency of
perturbations at higher redshift.  However, we see that this effect
does not appear to be result of the high density cluster enviroment:
Fig.~9b shows that $50\pm 8$\% of the cluster Sb--Sc--Sd/Irr galaxies have
$D\geq 2$, a proportion similar to that seen in the late-type field
population, $60\pm 11$\%.  The same effect is seen at low-redshift 
(Hashimoto \& Oemler 1999).

\subsection{Spectroscopic Properties 
of the Cluster Galaxies}

In Fig.~10 we quantify the distribution of morphological type for the
various spectral classes, for both cluster members {\it and} field
galaxies.  The strong, though broad, relation between morphology and
star-formation seen in low-redshift galaxies is present in this
intermediate redshift sample as well. Looking at the star-forming
population which causes the Butcher-Oemler effect we see a clear
tendency for these galaxies to be predominantly late-type systems
(Couch et al.\ 1994, 1998; Dressler et al.\ 1994; Oemler et al.\ 1997),
although here there is a tail of earlier-types (at least in the e(a)
and e(c) classes).   These active early-type (E and S0) galaxies
comprise a higher fraction of the field population than they do in  the
clusters. The two ``recently-active'' classes, k+a and a+k,
appear to have morphological distributions which are intermediate
between the passive and active cluster populations.  There seems to be
a clear distinction between k+a and a+k in the sense that the latter
are of later morphological type, though the small number of a+k types
limits the statistical certainty of this result.

%
%
\centerline{\psfig{file=f11.ps,angle=0,width=2.5in}
}

\noindent{\scriptsize
\addtolength{\baselineskip}{-3pt} 
\hspace*{0.3cm}
Fig.~11.\ The cumulative distribution of [O{\sc ii}]\,3727 EW
for three independent morphological bins for both cluster (solid line) and field (dotted line) populations. 

\addtolength{\baselineskip}{3pt}
}

It is interesting that, although the passive cluster population is
dominated by elliptical and S0 galaxies, there is a significant number
of later types, stretching out to Sd/Irr, which also show no emission
lines.  Aperture biases in our spectroscopy are unlikely to explain the
lack of observed star-formation in this group: the spectra sample the
central $\sim 65 h^{-2}$ kpc$^2$ of these distant galaxies.  Further
support for a lack of on-going star-formation in these systems is shown
by the uniform red colors of those galaxies for which we have imaging
in two passbands with {\it WFPC2}.  

We quantify the occurrence of passive late-type galaxies, and compare
cluster and field populations, in Fig.~11.  Using the cumulative
distribution of [O{\sc ii}]\,3727 EW, we find that for the
morphological groups E and S0--Sb there is a significantly higher
fraction of galaxies showing little or no [O{\sc ii}] emission in
clusters as compared to the field. The likelihood, $P$, that the
cluster and field samples are drawn from the same population is less
than $\log_{10} P <-2.4$ for both E and S0--Sb samples.  However, the
comparison of the [O{\sc ii}] distribution of the latest-type systems
(Sc--Irr, T=7--10) shows no significant difference between the cluster
and field, although the number of galaxies is somewhat smaller.

As an overall trend, then, there seems to be a decline in current
star-formation at a fixed Hubble-type from field to cluster (see also
Balogh et al.\ 1998).  Furthermore, based on [O{\sc ii}] EW alone as a
measure of star formation, we see no evidence for enhanced
star-formation in gas-rich cluster galaxies compared to the equivalent
morphological sample in the field. We discuss this incidence of passive
late-type galaxies further in P98.

%
%
\centerline{\psfig{file=f12.ps,angle=0,width=2.5in}
}

\noindent{\scriptsize
\addtolength{\baselineskip}{-3pt} 
\hspace*{0.3cm}
Fig.~12.\  A comparison of the distribution of D4000 measures in
the different morphological types, for both cluster and field galaxies.
 
\addtolength{\baselineskip}{3pt}
}

In contrast to these results based on [O{\sc ii}] EW, the distribution
of D4000 strengths (Fig.~12) is very similar for cluster and field:
the individual morphological types are indistinguishable in D4000 at
better than $\log_{10} P >-1$ in each case.  Thus, while [O{\sc ii}],
the tracer of current star-formation, shows a decline in the cluster,
this does not appear to be reflected in an index sensitive to the
star-formation averaged over a somewhat longer period of the recent
past ($\sim 1$--3 Gyrs).

\section{Results and Discussion}

\subsection{The Incidence of k+a/a+k and e(a) Galaxies}

Our spectral catalog exhibits one effect that is especially strong:
the incidence of k+a/a+k galaxies in distant clusters is very high
compared to the surrounding field.  Table~7 shows that in the cluster
sample we have 60 examples of k+a and 18 examples of a+k, totaling
18\% of the sample.  This is similar to the typical value of
$\sim$10--20\% found by magnitude-limited surveys of distant clusters
(DG92; Couch et al.\ 1998).  However, this value strongly contrasts
with the 7 occurrences, all k+a, found in the high-redshift field
sample, only 2\%.

Indeed, 4 of these 7 cases are either uncertain or border-line, a
far greater fraction than for the cluster sample, so an incidence of
$\sim$1\% is compatible with these data. For the low-redshift Las
Campanas Redshift Survey (hereafter LCRS), Zabludoff et al.\ (1996)
found an incidence of 0.2\%, but their selection criteria included a
stronger limit on H$\delta$ of 5.5 \AA\ and they note that the number
increases to 0.6\% when the limit is dropped to 4.5 \AA.  Hashimoto
(1998) has evaluated the occurrence of the spectral classes as defined
in this paper for the LCRS, and finds 2.3\% for the occurrence of
k+a/a+k types.  In summary, these data seem to point to at most a
factor two increase in the frequency of k+a/a+k types between the low- and 
intermediate-redshift field populations. This is in marked contrast to the
order-of-magnitude increase in the frequency of k+a types in rich
clusters.  At low redshift, this frequency is $\ls 1$\% (determined
using the Dressler and Shectman (1988) catalog), compared to the 18\%
found here for the $z \sim 0.5$ clusters.\footnote{Caldwell \& Rose
(1997) have reported a frequency of $\sim$15\% of notably stronger
Balmer lines in early type galaxies in five low-redshift clusters.
These are for the most part lower luminosity systems, with H$\delta <$
3.0 \AA, which the authors suggest are the remnants of earlier bursts.
The results of that study do not, therefore, conflict with the much
lower frequency found by Dressler and Shectman for stronger, more
luminous systems.}

Zabludoff et al.\ attributed many of the low-redshift field ``E+A's''
as due to mergers and strong interactions, since morphologies of this
type are often observed in the low-redshift examples.  The expected
evolution can be estimated from the change in the incidence of close
pairs (Zepf \& Koo 1989; Patton et al.\ 1997), which would be predicted
to be the parent population.  Patton et al.\ (1997) claim that the
proportion of close pairs (two galaxies within 20$h^{-1}$ kpc)
increases by a factor of $\sim 1.5$ between $z=0$ and $z=0.33$.
Extrapolating this behavior to $<\! z\! > =0.42$ would predict an
increase in the fraction of close pairs of $\sim 2$--3 over that seen
locally.  Although we see at most a factor of two increase in the k+a
population from low to high redshift using the LCRS and our sample,
this does not rule out that a significant fraction of field k+a's are
due to such mergers.  

Zabludoff et al.\ argue further that, as the merger/interaction
mechanism appears to be responsible for low-redshift field examples of
such galaxies, it is reasonable to conclude that mergers may also be
responsible for the k+a/a+k galaxies in the intermediate-redshift
clusters. However, the radically different evolution described above of
the k+a/a+k population between cluster and field environments suggests
that the cluster environment {\it is} crucial in either the formation
of {\it cluster} k+a/a+k galaxies, or in prolonging their visibility.
This could in part be due to an increased propensity for mergers in the
groups infalling into the intermediate-redshift clusters.  However, our
morphological analysis (S97) finds only a minority of cases of k+a
spectra where the galaxy shows signs of a classic two-body merger, as
Zabludoff et al. found for the low-redshift field examples.  We conclude,
then, that at least one mechanism other than mergers is responsible for 
the large fraction of k+a/a+k galaxies in intermediate-redshift clusters.

As we discuss in B98, the majority of k+a/a+k spectra are the result of
a sudden decline in star formation rate that followed a substantial
rise, or burst, of star-formation, leaving a population of A-stars to
dominate the light for $\sim10^9$ years.  Given the generic nature of
the star-formation history required to form an a+k/k+a, mergers are
obviously not a unique explanation for the k+a/a+k phenomena.  For
example, accretion of smaller satellites, instead of mergers of
comparable mass systems, is not inconsistent with the morphologies we
see. The greater fraction of a+k/a+k galaxies in the
intermediate-redshift clusters as compared to the field is likely to be
connected as well with the frequency of e(a) galaxies in these
environments (B98).

\subsection{The Distribution and Kinematic Properties of the Cluster Galaxies}

As a final exercise in the comparison of spectroscopic properties with
other cluster characteristics, we examine the radial distributions of
our cluster sample as a function of spectroscopic type.  We begin by
assigning field centers --- these positions are given in Table~2.
There is usually little ambiguity of this due to the presence of a D or
cD galaxy, these have been confirmed as the cluster centers in all
cases from our the weak lensing analysis in Smail et al.\ (1997b).
Even in more complex cases, such as Cl\,1447$+$26, the ambiguity in
choosing a center will play little role over the large range in radius
we investigate.

%
%
\centerline{\psfig{file=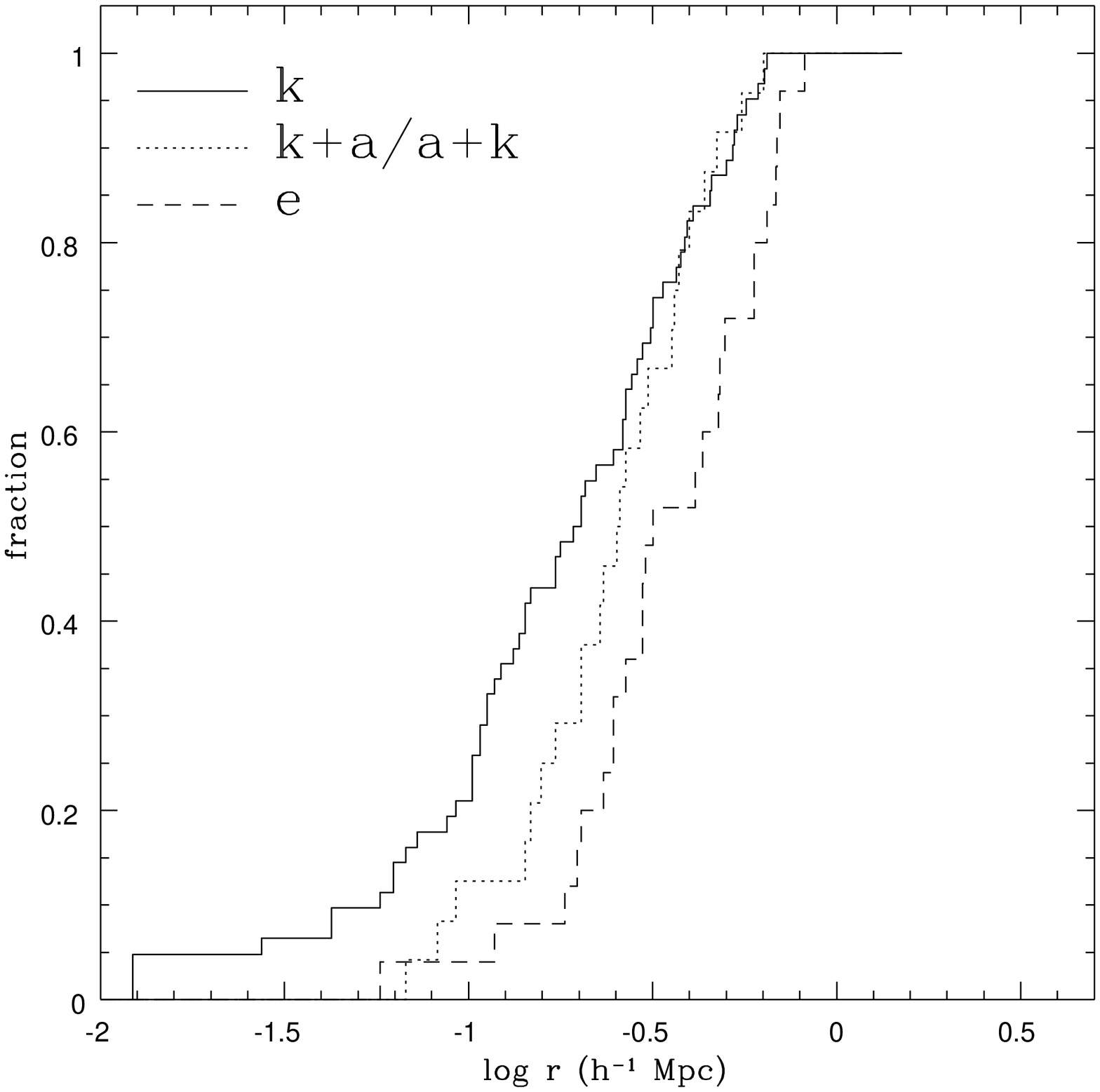,angle=0,width=2.5in}}

\noindent{\scriptsize
\addtolength{\baselineskip}{-3pt} 
\hspace*{0.3cm} Fig.~13.\ The cumulative radial
distribution of different spectral types.  These are shown for the all
members from the whole sample which have $M_V < -19 + 5 \log_{10} h$.
There is a clear difference between the radial distribution of k,
k+a/a+k, and e type galaxies, with the former being most concentrated,
the latter the least.  The k+a/a+k class seems to be intermediate
between the two showing a similar decline to the k types on the
outskirts of the cluster, but a flatter distribution in the core, more
in keeping with that seen for the e types.

\addtolength{\baselineskip}{3pt}
}

In Fig.~13 we show for the combined clusters the cumulative radial
distribution for different spectroscopic types.  This procedure is
crude because it averages over the non-spherical distribution of
galaxies within the clusters, but it may provide some insight into the
characteristic distributions of different classes of galaxies.  Not
surprisingly, the k types, generally made up of E and S0 galaxies
(Fig.~10), but including significant early-type spirals as well, are
the most concentrated population in these clusters (c.f.\ S97).  Also,
not surprisingly, the emission line galaxies strongly avoid the center
($r\leq 50 h^{-1}$\,kpc) of these clusters and have a much more
extended distribution.  What is perhaps more interesting is the way the
k+a/a+k types, which may sensibly interpreted as post-starburst
galaxies, avoid the centers in contrast to the k types, but are far
less extended than the emission-line galaxies.\footnote{It is tempting
to describe this distribution as a ``thick shell'', but we consider
this potentially misleading due to the substantial departures from
spherical symmetry exhibited by our clusters.  Rather, it is probably
more instructive to think of k+a/a+k types occurring most frequently at
an intermediate radius $R \sim 200 kpc$.} The near absence of k+a types
in the field, discussed above, coupled with the sudden rise in their
frequency as the cluster center is approached, with an almost complete
demise in the central regions, appears to be clear evidence for the
environment effecting either their formation or visibility.

We note in passing that a similar diagram subdividing the e types into
e(a), e(c), and e(b) shows no significant difference, though there is
a hint that the e(a) class has a slightly more extended distribution.

%
%
\begin{table*}
{\scriptsize
\begin{center}
\centerline{\sc Table 8}
\vspace{0.1cm}
\centerline{\sc Velocity Dispersions of Cluster Populations}
\vspace{0.3cm}
\begin{tabular}{lrrrr}
\hline\hline
\noalign{\smallskip}
 {Sample} & $N$~~ & {$<\! v\! >$~~} & {$\sigma$~~~~~} & {$\delta \sigma$~~~~~} \cr
~ & ~ & (km s$^{-1}$) &  (km s$^{-1}$)  & (km s$^{-1}$)  \cr
\hline
\noalign{\smallskip}
\multispan{5}{\hfil{Morphological~Types}\hfil}\cr
\noalign{\smallskip}
E       & 50 & $-$109.9 &  974.9 &  78.7 \cr
S0      & 24 & $-$378.7 & 1709.7 & 374.1 \cr
Sa     & 39 &     34.3 & 1336.5 & 137.2 \cr
Sb     & 38 &    179.5 & 1290.6 & 147.7 \cr
Sc     & 29 &    640.9 & 1212.6 & 219.2 \cr
Sd/Irr & 20 & $-$323.2 & 1894.6 & 231.9 \cr
\noalign{\medskip}
\multispan{5}{\hfil{Spectral~Classes}\hfil}\cr
\noalign{\smallskip}
k    & 186 &     12.5 & 1064.8 &  55.2 \cr
k+a  &  60 &    105.4 & 1421.6 & 204.9 \cr
a+k  &  18 &    120.9 & 1236.4 & 223.7 \cr
e(a) &  44 &  $-$27.2 & 1420.9 & 151.3 \cr
e(c) &  64 & $-$226.4 & 1437.1 & 126.2 \cr
e(b) &  20 & $-$168.8 & 1740.1 & 260.9 \cr
e(n) &   7 &    576.1 & 1495.3 & 404.9 \cr
e    &  13 &    743.2 & 1400.1 & 288.8 \cr
?    &  12 &    367.6 &  694.1 & 116.3 \cr
\noalign{\smallskip}
k      & 186 &    12.5  & 1064.8 &  55.2 \cr
k+a/a+k&  78 &   109.0  & 1373.4 & 172.4 \cr
e(all) & 148 & $-$36.3  & 1486.0 &  90.1 \cr
\noalign{\medskip}
\multispan{5}{\hfil{Disturbance~Classes}\hfil}\cr
\noalign{\smallskip}
$D=0$ & 100 &  $-$33.9 & 1252.9 & 144.9 \cr
$D=1$ &  54 &    207.6 & 1382.8 & 146.2 \cr
$D=2$ &  34 &     58.9 & 1417.5 & 202.3 \cr
$D=3$ &  12 &    307.5 & 1572.1 & 329.4 \cr
$D=4$ &   4 & $-$681.9 & 2884.3 & 807.8 \cr
\noalign{\medskip}
\multispan{5}{\hfil{Activity~Classes}\hfil}\cr
\noalign{\smallskip}
Sb-high & 10 &   ...   & 1429.5 & 294.1 \cr
Sb-low  & 28 &   ...   & 1265.5 & 173.6 \cr
Sc-high & 14 &   ...   & 1293.7 & 334.2 \cr
Sc-low  & 15 &   ...   & 1084.2 & 246.1 \cr
Sd-high & 10 &   ...   & 2274.5 & 332.7 \cr
Sd-low  & 10 &   ...   & 1243.1 & 238.1 \cr
\noalign{\smallskip}
\noalign{\hrule}
\noalign{\smallskip}
\end{tabular}
\end{center}
}
\vspace*{-0.8cm}
\end{table*}

%
%
\begin{figure*}
\centerline{\psfig{file=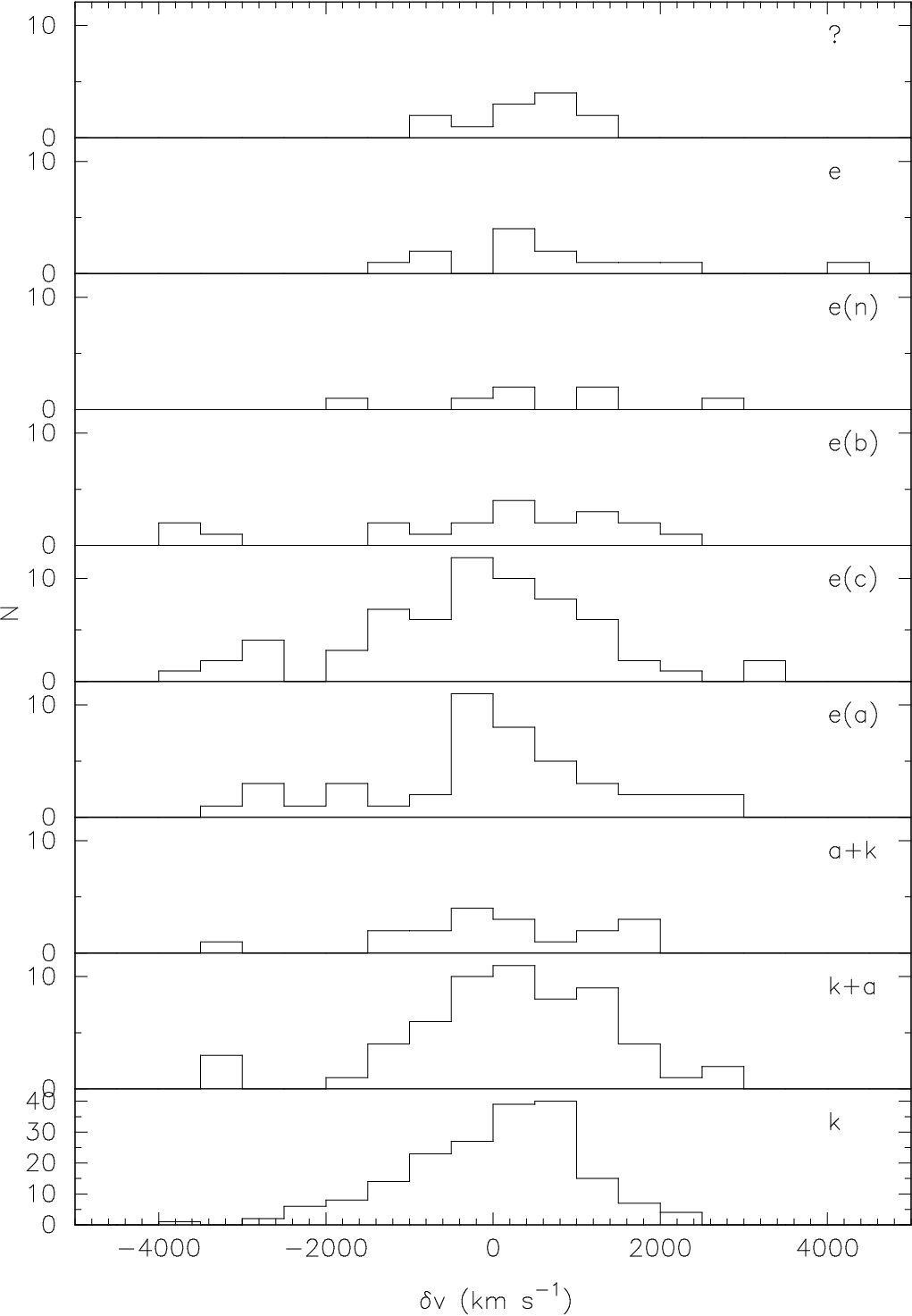,angle=0,width=2.5in}
\hspace*{0.5in}
\psfig{file=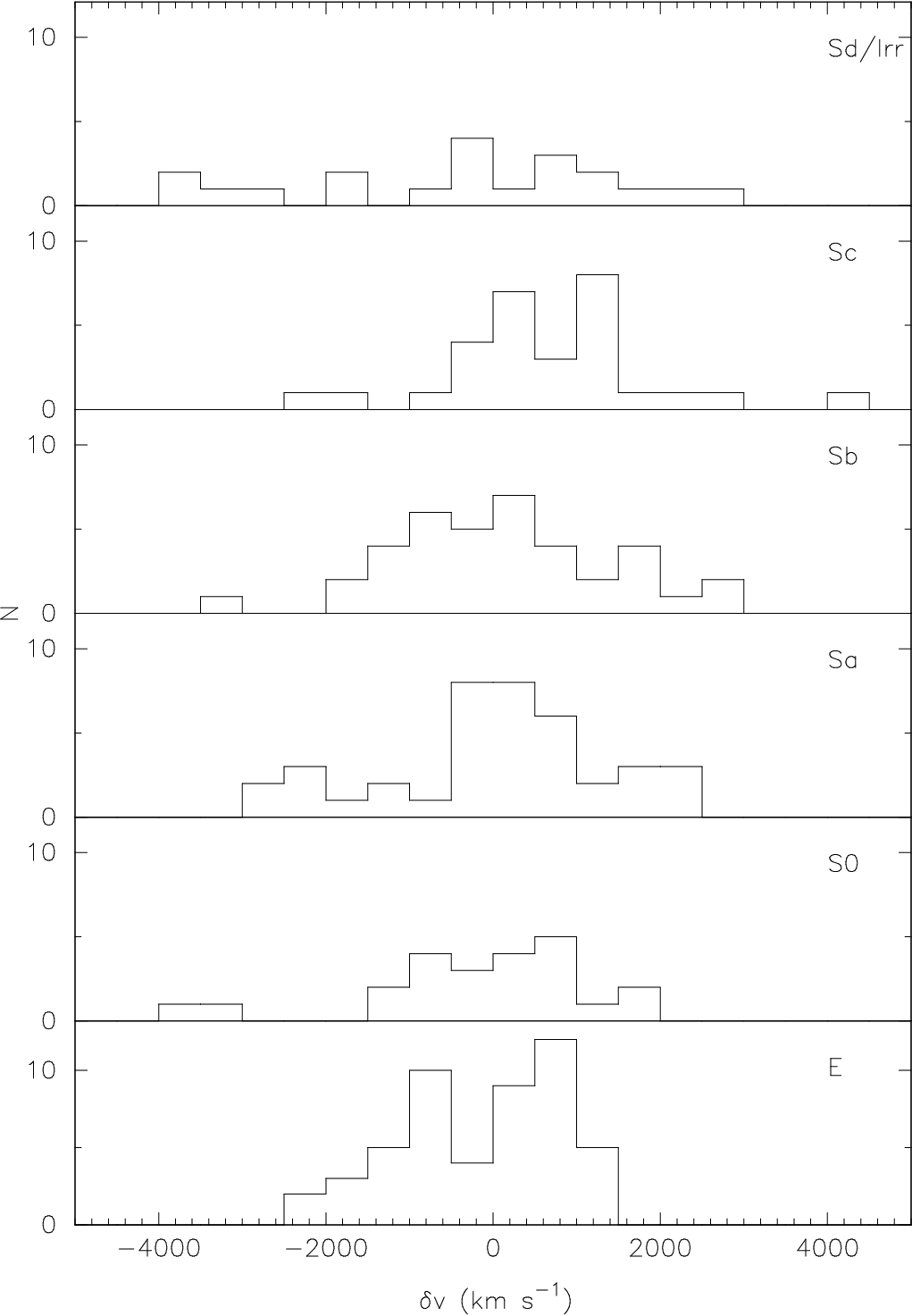,angle=0,width=2.5in}}
\noindent{\scriptsize
\addtolength{\baselineskip}{-3pt} 
\hspace*{0.3cm}
Fig.~14.\ a) The restframe velocities of cluster members, calculated
relative to their respective cluster means (Table~2).  These histograms
are shown for the different spectroscopic classes.  No scaling has been
applied when combining the distributions from different clusters.  b)
The restframe velocities of cluster members, relative to their
respective cluster means.  These histograms are shown for the different
morphological types.   

\addtolength{\baselineskip}{3pt}
}
\end{figure*}

We now investigate the rudimentary kinematics of the sample of cluster
members.  In Table~8 we list the restframe velocity dispersions and
uncertainties for the entire cluster sample broken down in terms of
morphological type, spectral class, disturbance and activity class (the
latter three sections refer only to those galaxies
lying within the {\it WFPC2} fields).  The distributions for the spectral and
morphological types are also shown in Fig.~14.  These values are
calculated using the mean cluster redshifts listed in Table~2 and are
simple averages across the cluster (no allowance has been made for
different velocity dispersion for the different clusters -- when such
corrections are applied they make no qualitative change to the
conclusions listed below).  The uncertainties in the velocity
dispersions are $1\sigma$ values estimated from bootstrap resampling of
the observed distributions.

Starting with the morphological samples in Table~8 we see a marked
difference between the velocity dispersion of the elliptical galaxies
and all the later-types, the latter having higher dispersions
(including the S0 galaxies).  A similar difference is noticeable when
the sample is split into different spectral classes (now including the
whole spectroscopic catalog of members).  Interestingly the galaxies
whose spectra were too poor to be classified, the ``?'' class, show the
lowest dispersion -- suggesting that these may be predominantly
passive, cluster galaxies.  

The strongest trend is the significantly higher velocity dispersion of
the presently or recently star-forming systems compared to the passive
population (c.f.\ Dressler 1986). In particular, combining the
different spectral classes (e(all) comprises e(a)/e(b)/e(c)/e(n)/e) 
from Table~8 we find that the emission-line and k-type galaxies have relative
dispersions of $\sigma_{em}/\sigma_k = 1.40\pm 0.16$, with the k+a/a+k
galaxies being intermediate between the two.  The higher dispersions of
the active populations are consistent with these galaxies being less
virialised than the k-type population.  Such a trend can also be
discerned in the variation of velocity with activity (as traced by the
[O{\sc ii}] EW) {\it within} the individual morphological types.
Splitting each of the more active morphological classes (Sb-Sd/Irr) 
at its median  [O{\sc ii}] EW into ``low'' and ``high'' activity
samples we find the dispersions listed at the bottom of Table~8
for the different morphological samples.
For all three morphological types, the more active sample
shows the higher velocity dispersion.  A higher velocity dispersion is
often taken as a sign of an infalling population, but, as we discuss
below, including spatial information in our analysis shows only
weak evidence for infall.

We note that the [O{\sc ii}] EW distributions for these active cluster
members do not show any enhanced activity over that seen in the
surrounding field for any given morphological type (c.f.\ \S4.2).
Apparently, then, the observed correlation of velocity dispersion and
activity (as measured by [O{\sc ii}] EW) is not triggered by a
mechanism which causes the higher star-formation rates due to the high
relative velocities of the galaxies within the clusters
(i.e.\ ram-pressure induced star-formation).  We suggest instead that
the correlation between activity and velocity dispersion reflects a
decline in star-formation in the galaxies which runs in parallel with
and is causally linked their virialization within the clusters.   
In this regard we also mention the trend for more
disturbed galaxies to have higher velocity dispersions, Table~8, a
result which remains when we restrict the analysis to late-type
galaxies, Sb--Irr.

We next combine the velocity and positional information for the entire
sample of clusters, focusing on the spectral classes, to calculate the
velocity dispersion as a function of spectral class and radius.
Dressler (1986) used the Giovanelli and Haynes (19ZZ) catalog of spirals in
nearby clusters to show that gas-poor spirals tend to travel on radial
orbits that take them into the cluster center, as compared to the more
isotropic orbits of the gas-rich spirals.  Here we divide the k,
k+a/a+k, and emission-line classes into three bins of radial position,
each containing one-third of their respective samples.  The resulting
velocity dispersions are shown in Fig.~15.  

%
%
\hbox{~}
\centerline{\psfig{file=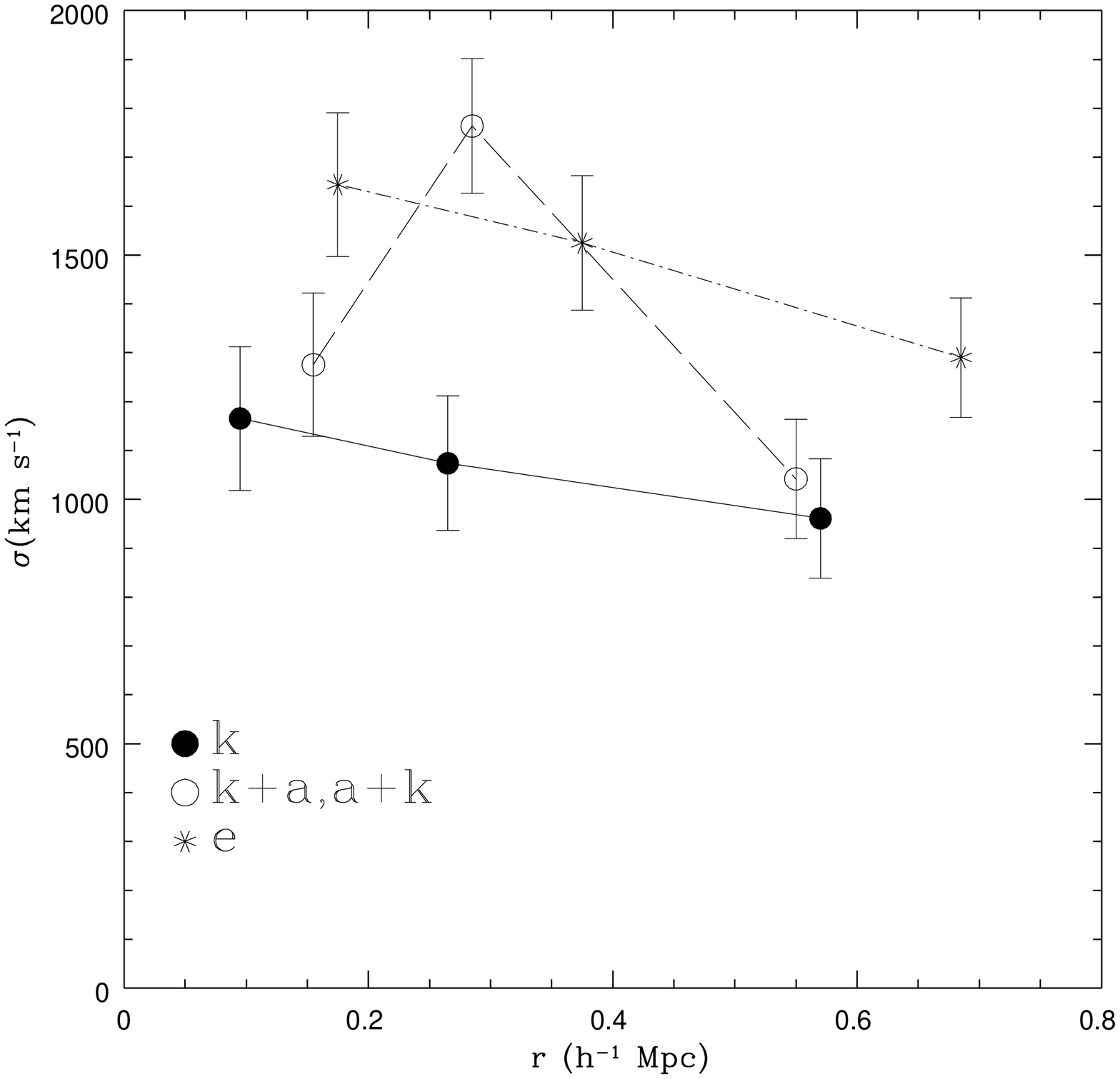,angle=0,width=2.5in}}
\noindent{\scriptsize
\addtolength{\baselineskip}{-3pt} 
\hspace*{0.3cm}
Fig.~15.\ The velocity dispersions of the different spectral types,
averaged over the entire sample, as a function of radius.  The velocity
dispersion is everywhere higher for active systems compared to passive
galaxies.    Note that the k+a/a+k types exhibit a  peak in velocity
dispersion that may be related to their distinctive spatial
distribution. 

\addtolength{\baselineskip}{3pt}
}

Unlike the clear difference in the orbital properties of gas-deficient
systems in nearby clusters, our sample exhibits ambiguous evidence at
best.  The k and e types both have velocity dispersion that falls
gently with radius or are, within the errors, flat.  This suggests
populations on mildly radial orbits, possibly an infalling population,
or a simple isothermal distribution. More puzzling is the peak in
velocity dispersion for the k+a/a+k types, which does appear to be
statistically significant.  It is possible, of course, that higher
velocities increase the chance of producing a k+a/a+k.  It is also
possible that this kinematic feature is connected with their unusual
radial distribution, as mentioned above. A system of largely circular
orbits that might characterize this distribution, which is concentrated
like the k-types but avoids the core, would appear to have a higher
velocity dispersion due to projection of what are largely tangential
velocities.  This is, however, not consistent with the idea that such
galaxies derive from an infalling population on what are basically
radial orbits.  At this point, the statistics are poor enough, and the
range of models so broad, that it is not worthwhile to explore this
further here.

\section{Conclusions}

\noindent{$\bullet$} We have presented detailed spectroscopic
observations of 657 galaxies in the fields of 10 $z=0.37$--0.56
clusters.  Combining these with our detailed {\it HST}-based
morphological catalogs in these fields we construct samples of 204
cluster members and 71 field galaxies with both accurate spectral and
morphological information.

\noindent{$\bullet$} Using observational and theoretical justifications
we have constructed a new quantitative spectral classification scheme
and use this to interpret correlations between our spectral information
and other properties of the galaxies in our catalog.

\noindent{$\bullet$} Based upon an analysis of the [O{\sc ii}] EW
distributions, we find no evidence for an increase in the occurrence of
strongly star-forming galaxies in the moderate-redshift cluster
environment compared to the moderate-redshift field using
morphologically-selected samples.  However, we do find a large
population of late-type cluster, but not field, galaxies which show
little or no evidence of on-going star-formation.

\noindent{$\bullet$} This passive, late-type cluster population is
related to our spectral classes k+a/a+k, both of which we interpret
as indicative of post-starburst behavior.  Galaxies with k+a/a+k
spectra are an order-of-magnitude more frequent in the cluster
environment compared to the high redshift field.  

\noindent{$\bullet$} These k+a/a+k galaxies avoid the central regions
of the clusters, in contrast to the k types, but are also far less
extended than the emission-line galaxies, and much less common in
the field.  This appears to be clear evidence for the environment
effecting either their formation or visibility.

\noindent{$\bullet$} A detailed analysis of the spectroscopic
and morphological information discussed here will be presented
in Poggianti et al.\ (1998).

\section*{Acknowledgements}

We thank Ray Lucas at STScI for his enthusiastic help which enabled the
efficient gathering of these HST observations.  BMP and HB warmly thank
Steve Maddox for crucial help during the 1995 WHT run and in the
subsequent reduction of those data. We also thank Alfonso
Arag\'on-Salamanca, Nobuo Arimoto and Amy Barger for useful discussions
and assistance.  AD and AO acknowledge support from NASA through STScI
grant 3857.  IRS acknowledges support from a PPARC Advanced Fellowship
and from Royal Society and Australian Research Grants while an Honorary
Visiting Fellow at UNSW. WJC acknowledges support from the Australian
Department of Industry, Science and Technology, the Australian Research
Council and Sun Microsystems.  This work was supported in part by the
Formation and Evolution of Galaxies network set up by the European
Commission under contract ERB FMRX-CT96-086 of its TMR program.  We
acknowledge the availability of the Kennicutt's (1992) atlas of
galaxies from the NDSS-DCA Astronomical Data Center.

\clearpage
\thispagestyle{empty}
%
%
\begin{sidetable}
{\scriptsize
\begin{center}
\centerline{\sc Table 4A\footnote{Complete versions of
Table 4A are available on AAS web site}}
\vspace{0.1cm}
\centerline{\sc Spectroscopic Catalog of Cluster Galaxies}
\vspace{0.3cm}
\hspace*{-1.0truein}
\begin{tabular}{lrlrrcclrrcrrlrclccccllll}
\hline\hline
\noalign{\smallskip}
Cluster & ID & ~~~$z$ & $q$~ & [O{\sc ii}] & H$\delta$ & 	D4000 & ~~Class & $\delta$RA~ & $\delta$Dec~ & ID$_{HST}$ & X~~ & Y~~ & ~Morph & T~ & D & Int &  Mag$^{a,b}$ & Col$^{a,c}$ & Mag$_{\rm DG}^d$ & Col$_{\rm DG}^d$ &    ~Run & ~~Mask & ~~~~Features & Comments \cr
~~(1) & (2) & ~~(3) & (4) & (5)~ & (6) & (7) & ~~~(8) & (9)~ & (10)~ & (11) & (12)~ & (13)~ & ~~(14) & (15) & (16) & (17) & (18) & (19) & (20) & (21) & ~(22) & ~~(23) & ~~~(24) & (25) \cr
\noalign{\smallskip}
\hline
\noalign{\smallskip}
0024   &    1  & 0.3910  &  4   &   0.0   & ...   & 2.52   & k:        &    69.1 &   $-$64.4 &   294 &  1317 &   343  & E          &  $-$5 &  0  & T?   &  19.96 &  3.30 &  21.54 &   1.66  & DG92   & ...       & ...                    &  DG16 \cr 
0024   &    2  & 0.3990  &  2   & $-$50.2   & ...   & 1.21   & e(b),CSB  &    97.4 &    11.8 &   871 &  1318 &  1365  & Scd        &   6 &  1  & I?   &  20.37 &  1.60 &  21.44 &   0.39  & DG92   & ...       & ...                    &  DG19; poss Hd str; H
b emis,OIII=$-$24 \cr 
0024   &    3  & 0.3825  &  3   &   0.0   & 2.3   & 2.13   & k         &    60.1 &   $-$15.3 &   739 &  1198 &  1008  & Sc         &   5 &  2  & I    &  20.12 &  2.56 &  21.01 &   1.10  & DG92   & ...       & ...                    &  DG59 \cr 
0024   &    4  & 0.3900  &  4   &   0.0   & ...   & 2.05   & k:        &    27.6 &   $-$15.2 &   573 &  1122 &   692  & E          &  $-$5 &  1  & M?   &  18.35 &  3.42 &  19.77 &   1.58  & DG92   & ...       & ...                    &  DG101 \cr 
0024   &    5  & 0.3910  &  2   & $-$14.4   & ...   & 1.28   & e(c)/e(a) &    42.3 &     7.2 &   659 &  1373 &   785  & Sd         &   7 &  1  & ...    &  19.97 &  1.95 &  21.30 &   0.76  & DG92   & ...       & ...                    &  DG104; Hb=$-$8.3,O
III=$-$2.9; poss Bal str -- can't measure Hd \cr 
0024   &    6  & 0.3900  &  2   &   0.0   & ...   & 1.56   & k         &    31.7 &    $-$5.5 &   577 &  1224 &   712  & S0         &  $-$2 &  0  & ...    &  19.51 &  3.18 &  20.86 &   1.26  & DG92   & ...       & ...                    &  DG106 \cr 
0024   &    7  & 0.3755  &  1   &   0.0   & ...   & 1.80   & k         &     0.9 &   $-$44.0 &   403 &   777 &   497  & S0/E       &  $-$3 &  0  & ...    &  19.35 &  3.32 &  20.66 &   1.45  & DG92   & ...       & ...                    &  DG112 \cr 
0024   &    8  & 0.3830  &  3   &   0.0   & ...   & 1.88   & k:        &    15.5 &   $-$14.4 &   461 &  1100 &   572  & S0/a       &   0 &  0  & I    &  20.08 &  2.88 &  21.49 &   1.38  & DG92   & ...       & ...                    &  DG120 \cr 
0024   &    9  & 0.3860  &  3   &   ...   & ...   & 1.85   & e(c):     &    12.0 &   $-$16.9 &   460 &  1069 &   543  & Sb         &   3 &  3  & I/M  &  19.84 &  3.32 &  21.45 &   1.47  & DG92   & ...       & ...                    &  DG123; wk OII, OIII 
emission? \cr 
0024   &   10  & 0.3950  &  1   &   ...   & ~3.6:  & 2.06   & k+a/k     &    46.7 &    38.6 &     ... &     ... &     ...  & ...          &   ... &  ...  & ...    &    ... &   ... &  20.26 &   1.48  & DG92   & ...       & H,K,G,Mg               &  DG128;Hd
 noisy \cr 
0024   &   11  & 0.3970  &  3   & $-$43.3   & 3.0   & 1.25   & e(b),CSB  &   $-$26.7 &   $-$46.9 &   203 &   689 &   234  & Sd         &   7 &  2  & T    &  20.32 &  1.49 &  21.48 &   0.37  & DG92   & ...       & ...                    &  DG138; Hb=$-$7.1
, OIII=$-$6.5 \cr 
0024   &   12  & 0.3960  &  3   &   0.0   & ...   & 2.01   & k:        &   $-$11.9 &   $-$24.2 &   304 &   942 &   327  & E          &  $-$5 &  0  & ...    &  18.47 &  3.42 &  19.63 &   1.58  & DG92   & ...       & ...                    &  DG140 \cr 
0024   &   13  & 0.3900  &  3   &   0.0   & ...   & 2.10   & k         &     4.9 &     2.8 &     ... &     ... &     ...  & ...          &   ... &  ...  & ...    &    ... &   ... &  19.29 &   1.61  & DG92   & ...       & ...                    &  DG144 \cr 
0024   &   14  & 0.3900  &  3   &   0.0   & ...   & 2.08   & k:        &     0.0 &     0.0 &   343 &  1204 &   387  & E          &  $-$5 &  0  & ...    &  18.35 &  3.51 &  19.49 &   1.71  & DG92   & ...       & ...                    &  DG148 \cr 
0024   &   15  & 0.3956  &  2   &   0.0   & ...   & 2.02   & k         &   $-$30.1 &   $-$22.7 &   133 &   914 &   146  & S0         &  $-$2 &  0  & ...    &  19.94 &  3.32 &  21.10 &   1.42  & DG92   & ...       & ...                    &  DG175 \cr 
0024   &   16  & 0.3990  &  2   & $-$20.7   & ...   & 1.49   & e(c):     &    23.9 &    53.6 &     ... &     ... &     ...  & ...          &   ... &  ...  & ...    &    ... &   ... &  21.34 &   0.84  & DG92   & ...       & ...                    &  DG177;
 Hb=$-$5.6,OIII*2 wk ($-$4.1); cannot meas Hd \cr 
0024   &   17  & 0.3902  &  1   &  $-$4.9   & 3.6   & 1.64   & e(c)      &   $-$30.9 &   $-$18.9 &   127 &   949 &   128  & Sbc        &   4 &  0  & ...    &  19.45 &  2.77 &  20.73 &   0.96  & DG92   & ...       & ...                    &  DG178 m,e \cr 
0024   &   18  & 0.3830  &  2   &   0.0   & ...   & 2.08   & k:        &   $-$15.6 &    33.5 &   147 &  1499 &   158  & E          &  $-$5 &  1  & I?   &  19.41 &  3.29 &  20.79 &   1.64  & DG92   & ...       & ...                    &  DG196 \cr 
0024   &   19  & 0.3940  &  1   & $-$137.3   & ...   & 0.81   & e(n),CSB  &   $-$70.2 &   $-$40.2 &     ... &     ... &     ...  & ...          &   ... &  ...  & ...    &    ... &   ... &  21.87 &   0.36  & DG92   & ...       & OII,Hb,OIII            &  D
G198; str emission -- Hb=$-$56.8, OIII=$-$124.8 \cr 
0024   &   20  & 0.3850  &  1   &   0.0   & ...   & 2.08   & k:        &   $-$18.9 &    39.3 &   113 &  1552 &   114  & E          &  $-$5 &  0  & I?   &  18.89 &  3.40 &  19.73 &   1.58  & DG92   & ...       & ...                    &  DG210 \cr
 .  & & .  & & .  & & .  & & .  & & .  & & .  & & .  & & .  & & .  & & .  & \cr
 .  & & .  & & .  & & .  & & .  & & .  & & .  & & .  & & .  & & .  & & .  & \cr
 .  & & .  & & .  & & .  & & .  & & .  & & .  & & .  & & .  & & .  & & .  & \cr
\noalign{\smallskip}
\noalign{\hrule}
\noalign{\smallskip}
\end{tabular}
\end{center}
}
\end{sidetable}

\clearpage
\thispagestyle{empty}
%
%
\begin{sidetable}
{\scriptsize
\begin{center}
\centerline{\sc Table 4B\footnote{Complete versions of
Table 4B are available on AAS web site}}
\vspace{0.1cm}
\centerline{\sc Spectroscopic Catalog of Field Galaxies}
\vspace{0.3cm}
\hspace*{-1.0truein}
\begin{tabular}{lrlrrcclrrcrrlrclccccllll}
\hline\hline
\noalign{\smallskip}
Cluster & ID & ~~~$z$ & $q$~ & [O{\sc ii}] & H$\delta$ & 	D4000 & ~~Class & $\delta$RA~ & $\delta$Dec~ & ID$_{HST}$ & X~~ & Y~~ & ~Morph & T~ & D & Int &  Mag$^{a,b}$ & Col$^{a,c}$ & Mag$_{\rm DG}^d$ & Col$_{\rm DG}^d$ &    ~Run & ~~Mask & ~~~~Features & Comments \cr
~~(1) & (2) & ~~(3) & (4) & (5)~ & (6) & (7) & ~~~(8) & (9)~ & (10)~ & (11) & (12)~ & (13)~ & ~~(14) & (15) & (16) & (17) & (18) & (19) & (20) & (21) & ~(22) & ~~(23) & ~~~(24) & (25) \cr
\noalign{\smallskip}
\hline
\noalign{\smallskip}
0024   &  108  & 0.1393  &  3   &   ...   & 3.1   & ...    & e(c)      &    28.9 &   $-$65.2 &   797 &  1221 &  1103  & Scd        &   6 &  0  &...   &  17.90 &  2.57 &  19.20 &   0.73  & DG92   & ...       & Ha,SII                 &  DG47; Ha=$-$25.5 Ha 
also includes [N II]? -- hi excit \cr 
0024   &  109  & 0.4545  &  3   & $-$12.9   & 0.0   & 1.41   & e(c),CSB  &   $-$28.4 &    43.6 &    ...&    ...&    ... &...         &  ...& ... &...   &    ... &   ... &  21.17 &   0.63  & DG92   & ...       & OII,OIII               &  DG220; hi excit \cr 
0024   &  110  & 0.2132  &  2   &   0.0   & 7.4   & 2.11   & e(a)      &   $-$76.7 &    57.6 &    ...&    ...&    ... &...         &  ...& ... &...   &    ... &   ... &  21.30 &   0.79  & P9408  & ew1\_11    & H,K,Bal,Ha             &  DG290; sum of ew1\_11 and ew2\_7; Bal abs and no OII, but Ha=$-$5.4 --- would be k+a if Ha not accessible \cr 
0024   &  111  & 0.2123  &  3   & $-$19.3   & 4.8   & 1.74   & e(a)      &   $-$35.4 &    $-$6.3 &    ...&    ...&    ... &...         &  ...& ... &...   &    ... &   ... &  21.97 &   0.84  & P9408  & ew1\_13    & OII,Hb,abs             &  DG185; also H\&
K and break; Ha$<$10 \cr 
0024   &  112  & 0.4758  &  2   & $-$43.8   & 6.5   & 1.54   & e(a)      &   168.3 &    66.7 &    ...&    ...&    ... &...         &  ...& ... &...   &    ... &   ... &  22.51 &   0.67  & P9408  & ew1\_27    & OII,Bal                &  DG13; also H,K,Mg \cr 
0024   &  113  & 0.6547: &  4   &   0.0   & 0.0   & 1.80   & k         &   228.3 &    43.6 &    ...&    ...&    ... &...         &  ...& ... &...   &    ... &   ... &    ... &    ...  & P9408  & ew1\_32    & H,K,G,Mg               &  noisy spectrum, but z
 may be correct \cr 
0024   &  114  & 0.2252  &  3   & $-$22.8   & 4.6   & 1.42   & e(a)      &   281.9 &   $-$44.8 &    ...&    ...&    ... &...         &  ...& ... &...   &    ... &   ... &    ... &    ...  & P9408  & ew1\_37    & OII,OIII,Ha            &  Ha=$-$29.0 \cr 
0024   &  115  & 0.2279  &  2   & $-$51.8   & 0.0   & 1.46   & e(b),CSB  &  $-$137.0 &    31.3 &    ...&    ...&    ... &...         &  ...& ... &...   &    ... &   ... &  23.31 &   0.17  & P9408  & ew1\_7     & OII,Hb,OIII,Ha         &  DG325; sum of ew1
\_7 and ew2\_4; v blue with mod str emission; Ha=$-$51.5 \cr 
0024   &  116  & 0.5558  &  2   & $-$27.2   & 4.7   & 1.35   & e(a)      &    $-$1.0 &    17.1 &   302 &  1371 &   341  & Scd        &  $-$5 &  2  & I?   &  20.18 &  2.82 &    ... &    ...  & P9412  & ew2\_15    & OII,OIII,Bal           &   \cr 
0024   &  117  & 0.7128  &  2   &   0.0   & 0.0   & 1.77   & k+a:        &   133.4 &    15.0 &    ...&    ...&    ... &...         &  ...& ... &...   &    ... &   ... &  22.20 &   1.52  & P9412  & ew2\_26    & H,K,G                  &  DG10 \cr 
0024   &  118  & 0.8089  &  4   & $-$15.9   & 0.0   & 1.55   & e(c)      &   177.0 &    45.7 &    ...&    ...&    ... &...         &  ...& ... &...   &    ... &   ... &    ... &    ...  & P9412  & ew2\_28    & OII,H,K                &  poor spectrum, but 
lines and continuum look right \cr 
0024   &  119  & 0.6569  &  2   &  $-$3.6   & 0.0   & 1.59   & e(c)      &   200.6 &    29.9 &    ...&    ...&    ... &...         &  ...& ... &...   &    ... &   ... &    ... &    ...  & P9412  & ew2\_29    & H,K,Mg,OII             &  weak OII emission in a K-type \cr 
0024   &  120  & 0.6564  &  2   &   0.0   & 0.0   & 1.92   & k         &   228.4 &    43.6 &    ...&    ...&    ... &...         &  ...& ... &...   &    ... &   ... &    ... &    ...  & P9412  & ew2\_30    & H,K,G,Mg               &   \cr 
0024   &  121  & 0.6946  &  2   & $-$37.8   & 3.9   & 1.22   & e(a)      &   246.1 &    32.9 &    ...&    ...&    ... &...         &  ...& ... &...   &    ... &   ... &    ... &    ...  & P9412  & ew2\_32    & OII,H,K,Mg,Bal         &  strong OII confirmed by wk Hb, OIII; \cr 
0024   &  122  & 0.1110  &  2   & $-$35.7   & 4.7   & 2.17   & e(a),CSB  &  $-$121.6 &    29.9 &    ...&    ...&    ... &...         &  ...& ... &...   &    ... &   ... &  21.44 &   0.29  & P9412  & ew2\_5     & OII,OIII,Ha            &  DG312 Ha$-$31.9 \cr 
0024   &  123  & 0.2153  &  1   &   0.0   & 0.0   & 1.85   & k         &   $-$64.7 &  $-$254.1 &    ...&    ...&    ... &...         &  ...& ... &...   &    ... &   ... &    ... &    ...  & P9410  & ns1\_1     & H,K,G,Hb,Mg            &  Hd$\sim$1 \cr 
0024   &  124  & 0.2478  &  2   & $-$53.4   & 2.3   & 1.63   & e(b)      &   $-$38.6 &   $-$58.7 &    ...&    ...&    ... &...         &  ...& ... &...   &    ... &   ... &    ... &    ...  & P9410  & ns1\_11    & OII,Hb,OIII,Ha         &  str emission lines on red continuum Ha=$-$38.4 \cr 
0024   &  125  & 0.1111  &  2   & $-$46.0   & 5.0   & 1.49   & e(b),e(a) &  $-$121.6 &    29.8 &    ...&    ...&    ... &...         &  ...& ... &...   &    ... &   ... &    ... &    ...  & P9410  & ns1\_19    & OII,Hb,OIII,Ha         &  Ha=$-$29.3 \cr 
0024   &  126  & 0.2129  &  1   & $-$29.5   & 1.3   & 1.81   & e(c)      &   $-$57.4 &    38.9 &    ...&    ...&    ... &...         &  ...& ... &...   &    ... &   ... &  19.31 &   0.59  & P9410  & ns1\_20    & OII,Hb,OIII,Ha         &  DG257; also hi ord Bal, metal ab lines Ha=$-$41.1 (including [N II]); Hd well measured \cr 
0024   &  127  & 0.2724  &  1   &   0.0   & 0.0   & 1.93   & e(c):     &   $-$53.9 &    86.8 &    ...&    ...&    ... &...         &  ...& ... &...   &    ... &   ... &  20.40 &   1.27  & P9410  & ns1\_24    & H,K,G                  &  DG289; rather wk abs; broad Ha/[N II] em; Hd$<$1 \cr 
 .  & & .  & & .  & & .  & & .  & & .  & & .  & & .  & & .  & & .  & & .  & \cr
 .  & & .  & & .  & & .  & & .  & & .  & & .  & & .  & & .  & & .  & & .  & \cr
 .  & & .  & & .  & & .  & & .  & & .  & & .  & & .  & & .  & & .  & & .  & \cr
\noalign{\smallskip}
\noalign{\hrule}
\noalign{\smallskip}
\end{tabular}
\end{center}
}
\end{sidetable}
\clearpage


%
%
{Fig.~8a.}  The  images of those galaxies in our sample lying within
the  {\it WFPC2} frame, grouped into spectroscopic classes.  This first panel
shows those galaxies with k spectral types.  Each image is $5''\times
5''$ (or 15.5--18.6 $h^{-1}$\, kpc depending upon the cluster's
redshift) and has the same orientation as the {\it HST} field (S97).
The cluster and galaxy ID from the {\it WFPC2} catalogs (from Tables~4 in
S97, or Table~4a below) and the spectral class and morphological type
are marked on each frame.  For the plates in Fig.~8 see http://www.ociw.edu/$\sim$irs/morphs2.html\#figs
\medskip

\centerline{{Fig.~8a.}  continued.} 
\medskip

\centerline{{Fig.~8a.}  continued.} 
\medskip

\centerline{{Fig.~8b.}  The k+a sample.}
\medskip

\centerline{{Fig.~8c.}  The a+k sample.}
\medskip

\centerline{{Fig.~8d.}  The e(a) sample.}
\medskip

\centerline{{Fig.~8e.}  The e(c) sample.}
\medskip

\centerline{{Fig.~8f.}  The e(b) sample.}
\medskip

\centerline{{Fig.~8g.}  The e(n) sample.}
\medskip


\end{document}